\documentclass[journal,10pt]{IEEEtran}
\usepackage{cite}
\usepackage{graphicx,color,epsfig,rotating}
\usepackage{amsfonts,amsmath,amssymb,bbm}
\usepackage{algorithm}
\usepackage{algpseudocode}
\usepackage{subfigure}
\usepackage{amsmath}
\usepackage{cite}
\usepackage{placeins}
\usepackage{graphicx}
\usepackage[latin1]{inputenc}
\usepackage{amssymb}
\usepackage{multirow}
\usepackage{stfloats}
\usepackage{tabularx} 
\usepackage{booktabs} 
\usepackage{url}
\usepackage{bm}
\usepackage{soul}
\usepackage{float}
\usepackage{lipsum}     
\usepackage{cuted} 
\usepackage{pstricks}
\usepackage{arydshln}
\usepackage{xspace} 

\allowdisplaybreaks

\long\def\comment#1{}

\makeatletter

\newcommand{\thickhline}{%
    \noalign {\ifnum 0=`}\fi \hrule height 1pt
    \futurelet \reserved@a \@xhline
}
\newcolumntype{"}{@{\hskip\tabcolsep\vrule width 1pt\hskip\tabcolsep}}
\makeatother

\makeatletter
\newcommand{\subalign}[1]{
  \vcenter{%
    \Let@ \restore@math@cr \default@tag
    \baselineskip\fontdimen10 \scriptfont\tw@
    \advance\baselineskip\fontdimen12 \scriptfont\tw@
    \lineskip\thr@@\fontdimen8 \scriptfont\thr@@
    \lineskiplimit\lineskip
    \ialign{\hfil$\m@th\scriptstyle##$&$\m@th\scriptstyle{}##$\crcr
      #1\crcr
    }%
  }
}
\makeatother

\newcommand{\re}[1]{{\color{red} #1}}

\newtheorem{example}{Example}
\newtheorem{theorem}{Theorem}
\newtheorem{lemma}{Lemma}
\newtheorem{corollary}{Corollary}
\newtheorem{remark}{Remark}

\def \Kexclk{{[K] \bksl  \{k\}}}
\def \inputsum{{\sum_{i=1}^KW_i}}
\def \rzsigma{{R_{Z_{\Sigma}}}}
\def \rzsigmastar{{R_{Z_{\Sigma}}^*}}
\def \rx{{R_X}} 

\def \rz{{R_Z}} 
\def \lx{{L_X}} 
\def \lz{{L_Z}}

\def  \zsigma{{Z_{\Sigma}}}
\def  \lzsigma{{L_{Z_{\Sigma}}}}

\let\trm\textrm
\let\tbf\textbf
\let\tit\textit

\let\mbb\mathbb

\let \bksl\backslash
\let \ovst \overset

\newcommand{\Iwf}{In what follows\xspace}

\newcommand{\Thf}{Therefore\xspace}

\newcommand{\Aar}{As a result\xspace}
\newcommand{\insum}{input sum\xspace}

\newcommand{\brdcst}{broadcast\xspace}

\newcommand{\corrspdg}{corresponding\xspace}
\newcommand{\dsa}{decentralized secure aggregation\xspace}
\newcommand{\Dsa}{Decentralized secure aggregation\xspace}

\newcommand{\decen}{decentralized\xspace}

\newcommand{\skr}{source key rate\xspace}

\newcommand{\eg}{e.g.\xspace}

\newcommand{\ie}{i.e.\xspace}
\newcommand{\msg}{message\xspace}
\newcommand{\msgs}{messages\xspace}
\newcommand{\Wlog}{Without loss of generality\xspace}

\newcommand{\hie}{hierarchical\xspace}

\newcommand{\Msp}{More specifically\xspace}

\newcommand{\Ip}{In particular\xspace}
\newcommand{\af}{as follows\xspace}
\newcommand{\resp}{respectively\xspace}
\newcommand{\iid}{i.i.d.\xspace}

\newcommand{\Thm}{Theorem\xspace}
\newcommand{\schm}{scheme\xspace}

\newcommand{\info}{information\xspace}

\newcommand{\itic}{information-theoretic\xspace}

\newcommand{\etal}{\textit{et al.}\xspace}

\newcommand{\Ovhd}{Overhead\xspace}

\newcommand{\agg}{aggregation\xspace}
\newcommand{\aggtr}{aggregator\xspace}
\newcommand{\secagg}{secure aggregation\xspace}
\newcommand{\Secagg}{Secure aggregation\xspace}
\newcommand{\SecAgg}{Secure Aggregation\xspace}
\newcommand{\Agg}{Aggregation\xspace}

\newcommand{\diff}{different\xspace}
\newcommand{\Diff}{Different\xspace}

\newcommand{\indep}{independent\xspace}

\newcommand{\indepce}{independence\xspace}
\newcommand{\indeply}{independently\xspace}

\newcommand{\indiv}{individual\xspace}
\newcommand{\Indiv}{Individual\xspace}
\newcommand{\comm}{communication\xspace}
\newcommand{\Comm}{Communication\xspace}

\newcommand{\achvs}{achieves\xspace}
\newcommand{\achved}{achieved\xspace}
\newcommand{\achvb}{achievable\xspace}

\newcommand{\distn}{distribution\xspace}

\newcommand{\muinfo}{mutual information\xspace}

\newcommand{\Feon}{For ease of notation\xspace}

\newfont{\bbb}{msbm10 scaled 700}

\newfont{\bb}{msbm10 scaled 1100}

\newcommand{\kth}{{$k^{\rm th}$ }}

\newcommand{\Cc}{{\cal C}}

\newcommand{\Rc}{{\cal R}}
\newcommand{\Sc}{{\cal S}}
\newcommand{\Tc}{{\cal T}}

\newcommand{\Xc}{{\cal X}}
\newcommand{\Yc}{{\cal Y}}

\newcommand{\eqdef}{\stackrel{\Delta}{=}}

\newcommand{\be}{\begin{equation}}
\newcommand{\ee}{\end{equation}}
\newcommand{\bea}{\begin{eqnarray}}
\newcommand{\eea}{\end{eqnarray}}

\begin{document}

\title{Information-Theoretic Decentralized Secure Aggregation  with Passive Collusion Resilience}

\author{
Xiang~Zhang,~\IEEEmembership{Member,~IEEE},
Zhou Li,~\IEEEmembership{Member,~IEEE},
Shuangyang Li,~\IEEEmembership{Member,~IEEE},
Kai Wan,~\IEEEmembership{Member,~IEEE},
Derrick Wing Kwan Ng,~\IEEEmembership{Fellow,~IEEE},
and Giuseppe Caire,~\IEEEmembership{Fellow,~IEEE}

\thanks{X. Zhang, S. Li, and G. Caire are with the Department of Electrical Engineering and Computer Science, Technical University of Berlin, 10623 Berlin, Germany (e-mail: \{xiang.zhang, shuangyang.li,caire\}@tu-berlin.de).
}

\thanks{Z. Li is the Guangxi Key Laboratory of Multimedia Communications and Network Technology, Guangxi University, Nanning 530004, China (e-mail: lizhou@gxu.edu.cn).}

\thanks{
K. Wan is with the School of Electronic Information and Communications,
Huazhong University of Science and Technology, Wuhan 430074, China
(e-mail: kai\_wan@hust.edu.cn).}

\thanks{
D. W. K. Ng is with the School of Electrical Engineering and Telecommunications, University of New South Wales, Sydney, NSW 2052, Australia
(e-mail: w.k.ng@unsw.edu.au).}

}

\maketitle

\begin{abstract} 
In decentralized federated learning (FL), multiple clients collaboratively learn a shared machine learning (ML) model by leveraging their privately held datasets distributed across the network, through interactive exchange of the intermediate model updates.
To ensure data security, cryptographic techniques are commonly employed to protect model updates
during aggregation.
Despite growing interest in secure aggregation, existing works predominantly focus on protocol design and computational guarantees, with limited understanding of the fundamental information-theoretic limits of such systems. Moreover, optimal bounds on communication and key usage remain unknown in decentralized settings, where no central aggregator is available. 
Motivated by these gaps, we study the problem of decentralized secure aggregation (DSA) from an information-theoretic perspective. Specifically, we consider a network of $K$ fully-connected users, each holding a private input---an abstraction of local training data---who aim to securely compute the sum of all inputs. The security constraint requires that no user learns anything beyond the input sum, even when colluding with up to $T$ other users.
We characterize the optimal rate region, which specifies the minimum achievable communication and secret key rates for DSA. In particular, we show that to securely compute one symbol of the desired input sum, each user must (i) transmit at least one symbol to others, (ii) hold at least one symbol of secret key, and (iii) all users must collectively hold no fewer than  $K - 1$ independent key symbols.
Our results establish the fundamental performance limits of DSA, providing insights for the design of provably secure and communication-efficient protocols in decentralized learning.
\end{abstract}


\section{Introduction}
\label{sec: intro} 
Federated learning (FL) has emerged as a powerful paradigm for distributed machine learning (ML), enabling users to collaboratively train a global ML model without directly sharing their private datasets~\cite{mcmahan2017communication,konecny2016federated,kairouz2021advances,yang2018applied}. 
Since each user's local dataset resides exclusively on their own device and is not shared with the \agg server or other users, FL inherently offers enhanced data security and user privacy. A typical FL training process includes two phases, a local training phase followed by a global \agg phase~\cite{mcmahan2017communication,karimireddy2020scaffold}.
In the local training phase, multiple stochastic gradient descent (SGD)-based optimization steps~\cite{kingma2014adam,reddi2019convergence} are performed independently on each user's private dataset to optimize a local loss function. After local training, the users upload their model parameters to a central server, which computes an aggregated model---typically a weighted sum of the local models. The aggregated model is then sent back to the users, serving as a starting point for a new round of local training. The training process alternates between the local training and global \agg phases, enabling timely correction of gradient divergence arising from non-identically distributed local datasets~\cite{li2019convergence}.

More recently, FL has been studied in a serverless 
setting known as \emph{decentralized} FL (DFL)~\cite{he2018cola}, where the aggregation of model parameters is distributed across neighboring nodes over a communication graph, thus enabling fast local updates.
Compared to conventional centralized FL, DFL offers several benefits~\cite{beltran2023decentralized}. First, it improves fault tolerance by allowing nodes to dynamically adapt to changes in network topology, avoiding single points of failure in the centralized case~\cite{wang2021edge}. Second, it alleviates network bottlenecks by enabling more evenly distributed communication and computation workload across the participating users~\cite{savazzi2021opportunities}.

Although private local datasets are never directly shared with other users or the server, FL remains vulnerable to certain security attacks~\cite{bouacida2021vulnerabilities, geiping2020inverting, mothukuri2021survey,dontu2024attack}. For example, it was shown that a substantial amount of information from a user's image dataset can be inferred by the aggregation server via model inversion attacks~\cite{geiping2020inverting}, simply by observing the sequence of uploaded model updates. As a result, additional security measures are required to ensure security in FL, motivating the development of {\secagg (SA) protocols~\cite{bonawitz2017practical, bonawitz2016practical, 9834981,wei2020federated,hu2020personalized,zhao2020local,yemini2023robust,so2021turbo, liu2022efficient,jahani2023swiftagg+,seranmadevi2026security} in FL}. 
In a pioneering work, Bonawitz \etal~\cite{bonawitz2017practical} proposed a secure aggregation scheme based on pairwise random seed agreements between users, which are used to generate zero-sum masks that conceal each user's local model from the server. Since then, various cryptographic techniques based on random seed agreement have been used  to achieve \emph{computational} security.

In a different vein, \itic \secagg~\cite{9834981,zhao2023secure} investigates the fundamental performance limits of \secagg under the lens of \emph{perfect} security, characterized by a zero mutual information criterion. In this \itic  formulation, the locally trained models of the users are abstracted as \iid \emph{inputs}, denoted by $W_1,\cdots, W_K$, where $K$ is the total number of users. 
The \agg server wants to compute the sum of inputs $W_1 +\cdots+W_K$, subject to the security  constraint that the server should not infer any \info about the input set $(W_1,\cdots,W_k)$ beyond their sum. Formally, the security constraint is expressed as 
\be 
\label{eq:security,I=0,intro}
I\left(\{X_k\}_{k=1}^K; \{W_k\}_{k=1}^K|W_1 +\cdots +W_K\right)=0,
\ee
where $X_k$  is the message sent from User $k\in[K]$. This condition ensures that the messages observed by the server are statistically \indep of the inputs, thereby achieving perfect security. It should be noted that random seed-based cryptography cannot satisfy (\ref{eq:security,I=0,intro}) due to Shannon's one-time pad (OTP) encryption theorem~\cite{shannon1949communication}.  
Although \comm  overhead has been extensively studied in SA, the optimization of randomness consumption for key generation has received comparatively little attention. 
In contrast, \itic SA addresses both the communication and key generation aspects of the \agg process, aiming to characterize the optimal rate region---that is, the set of communication and key rate pairs that can be simultaneously achieved.

Recent studies on information-theoretic SA have extended its scope to a wide range of practical constraints, including user dropout and collusion resilience~\cite{9834981, zhao2023secure,so2022lightsecagg,jahani2022swiftagg,jahani2023swiftagg+}, groupwise keys~\cite{zhao2023secure,wan2024information,wan2024capacity}, user selection~\cite{zhao2022mds, zhao2023optimal}, heterogeneous security~\cite{li2023weakly,li2025weakly,li2025collusionresilienthierarchicalsecureaggregation}, oblivious server~\cite{sun2023secure}, SA with multiple recovery goals~\cite{yuan2025vector}, and \hie secure \agg (HSA)~\cite{zhang2024optimal, 10806947,egger2024privateaggregationhierarchicalwireless, lu2024capacity,zhang2025fundamental,li2025collusionresilienthierarchicalsecureaggregation}. \Msp,
Zhao and Sun~\cite{9834981} studied SA in a server-based star network, accounting for the effects of both user dropout (straggling users) and server collusion.   
Secure \agg schemes with improved key storage and \distn overhead were studied in~\cite{ so2022lightsecagg,so2021turbo}.
SA with uncoded groupwise keys was explored in~\cite{zhao2023secure,wan2024information,wan2024capacity}, where each subset of users of a given size shares an independent key. The motivation behind using groupwise keys is that they can be efficiently generated via interactive key agreement protocols among users, thereby eliminating the need for a designated key distribution server. Moreover, SA with heterogeneous security constraints across users-referred to as weak security---was investigated in~\cite{li2023weakly,li2025weakly,li2025collusionresilienthierarchicalsecureaggregation}. 
SA with user selection was studied in~\cite{zhao2022mds, zhao2023optimal}, where the server aims to recover the sum of inputs from a predefined subset user, capturing the effect of partial user participation in practical FL systems. 
{In addition, privacy-preserving distributed consensus has also been studied in \cite{li2019privacy,li2021privacy}, which is relevant to \secagg but focus more on the algorithmic aspects.}

It can be seen that existing studies on information-theoretic secure aggregation have predominantly focused on the centralized server-client setting, sometimes incorporating an intermediate relay layer. Motivated by the limited understanding of \itic SA in \decen  FL, this paper investigates the fundamental limits of \emph{\dsa} (DSA) under a $K$-user fully-connected network topology\footnote{This network model has been  widely adopted in the literature, such as device-to-device (D2D) coded caching~\cite{ji2015fundamental} and coded distributed computing (CDC)~\cite{li2017fundamental}.}. In particular, each user possesses a private input $W_k$ and is connected to all other users through an error-free broadcast channel as shown in Fig.~\ref{fig:model}. 
The broadcast channels originating from different users are assumed to be orthogonal, allowing simultaneous and interference-free transmissions.
Each user aims to compute the sum of inputs of all users $W_1+\cdots+W_K$, while satisfying the \emph{security constraint} that no user should infer any \info about the input set  $(W_1, \cdots, W_K)$ beyond their sum and what can be learned through collusion with at most $T\le K-1$ other users. To achieve security, each user is also equipped with a key variable $Z_k$, based on which a broadcast message $X_k$ can be generated. Assume the \indiv keys $Z_1, \cdots,Z_K$ are derived from a source key variable $\zsigma$, \ie, $H(Z_1,\cdots,Z_K|\zsigma)=0$.
Let $R_X, R_Z$ and $\rzsigma$ denote the per-user \comm rate, \indiv key rate, and the source key rate, which \resp quantifies the sizes (normalized by the input length)  of $X_k,Z_k$ and $\zsigma$. 
We aim  to find the optimal rate region $\Rc^*$, which consists of all simultaneously achievable rate triples $(R_X, R_Z, \rzsigma)$.

The core contribution of this work is a complete characterization of the optimal rate region $\Rc^*$ of DSA. Specifically, we show that $\Rc^*$ is empty when $T\ge K-2$,  and equal to    $\{(R_X, R_Z, \rzsigma):\rx\ge 1, \rz\ge 1, \rzsigma \ge K-1\}$ when $K\ge 3$ and $T\le K-3$.
This reveals a fundamental limit: \emph{to securely compute one symbol of the input sum, each user must transmit at least one symbol, hold at least one symbol of secret key, and the system must collectively hold at least $K - 1$ independent key symbols.}
This characterization is achieved via a linear \achvb scheme, paired with a set of entropic converse bounds that tightly establish the minimum \comm and key rates. Together, they provide the first information-theoretic characterization of the fundamental limits of secure aggregation in fully decentralized settings.

\subsection{Summary of Contributions}
To summarize, the contributions of this paper include: 
\begin{itemize}
    \item An \itic formulation of the \dsa problem, including a rigorous definition of the security constraint based on mutual information.

    \item A linear and optimal aggregation scheme that ensures both correct recovery and security for all users, even in the presence of collusion. The scheme achieves the minimum per-user \comm, \indiv key, and source key rates, thereby maximizing both communication and key generation efficiency.

    \item  A novel converse proof establishing tight lower bounds on the achievable rate region. The matching bounds certify the optimality of the proposed scheme. Furthermore, the converse techniques developed in this work are broadly applicable and may extend to decentralized settings with more general (non-fully-connected) network topologies.
\end{itemize}

\tit{Paper Organization.}
The remainder of this paper is organized as follows.
Section~\ref{sec: problem description} introduces the problem formulation.
Section~\ref{sec:main result} presents the main result along with its theoretical implications.
Section~\ref{sec: ach scheme} describes the proposed secure aggregation scheme, with design insights illustrated through a motivating example.
Section~\ref{sec: converse} provides the general converse proof.
Finally, Section~\ref{sec:conclusion & future directions} concludes the paper with a discussion of future research directions.

\tit{Notation.} Throughout the paper, the following  notations are used. For integers $m\le n$, let $[m:n] \eqdef \{m,m+1, \cdots, n\}$ and $[n]\eqdef\{1,2,\cdots,n\}$. $[m:n]=\emptyset$ if $m>n$.
For set of random variables $X_1,\cdots, X_N$, denote $X_{1:N}\eqdef\{X_1,\cdots, X_N\}$. Given two sets $ \Xc$  and $\Yc$, the difference set is defined as $\Xc \bksl \Yc\eqdef \{x\in \Xc: x\notin \Yc\}$. 
$H(X)$ denotes the entropy of $X$. $I(X;Y)$ denotes the mutual \info between $X$ and $Y$.

\section{Problem Statement}
\label{sec: problem description}
\begin{figure}[t]
    \centering
    \includegraphics[width=0.45\textwidth]{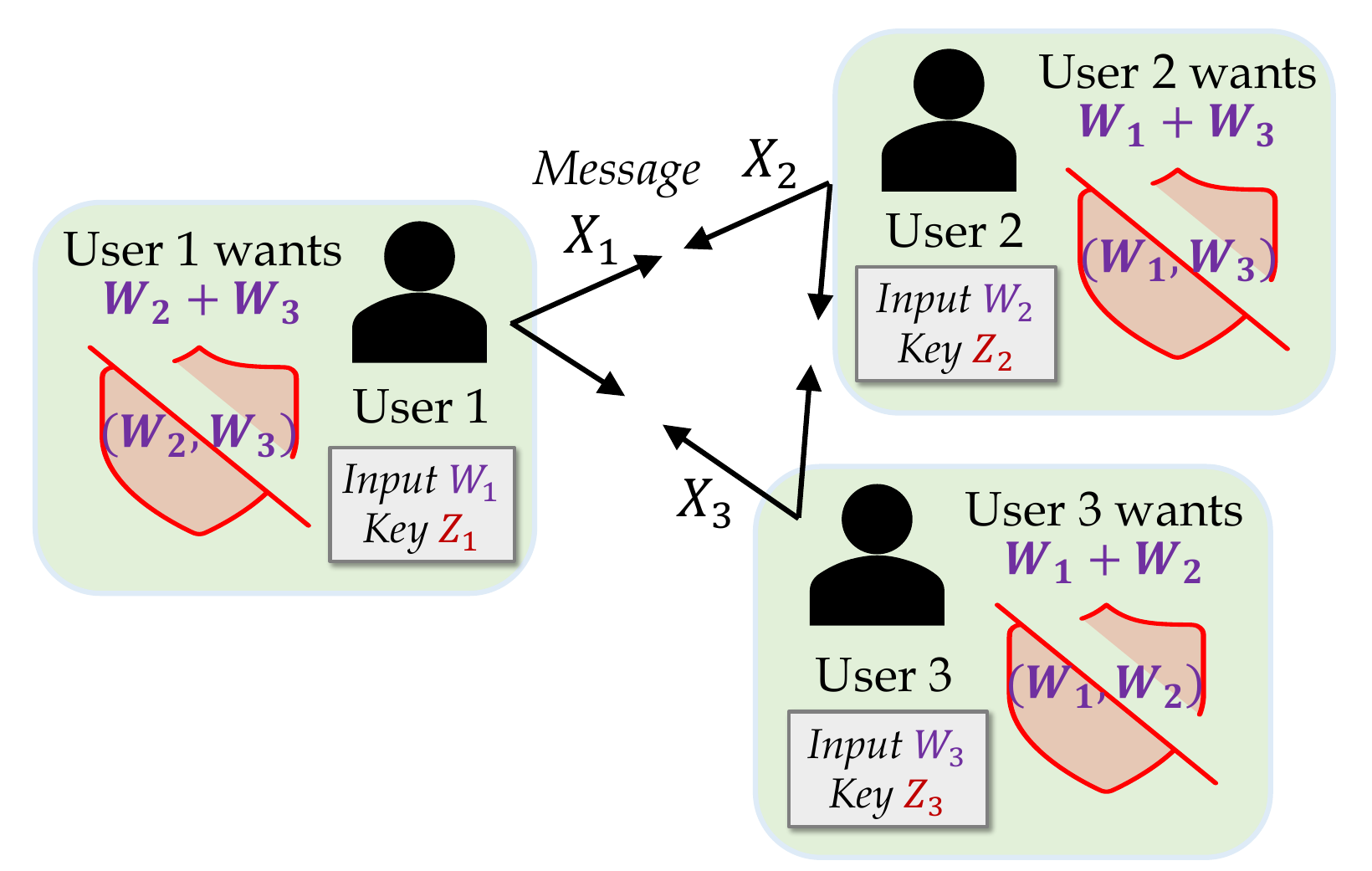}
    \vspace{-.2cm}
    \caption{\small \Dsa with 3 users. User 1 aims to recover the sum $W_2 + W_3$ from the received messages $X_2$ and $X_3$, while being prevented from learning any additional information about the pair of inputs $(W_2, W_3)$ beyond their sum. The same security requirement is imposed on the other two users as well.  }
    \label{fig:model}
    \vspace{-.1cm}
\end{figure}

Consider a system of $K\ge 3$ users, where each user is connected to the remaining users through an error-free broadcast channel as shown in Fig.~\ref{fig:model}.
Each user $k \in [K]$ holds an \emph{input} $W_k$, consisting of $L$ \iid uniformly distributed symbols over some finite field $\mathbb{F}_q$\footnote{{The uniformity assumption of the inputs is made solely for the purpose of the converse proof, \ie,  to derive lower bounds on the \comm and secret key rates defined later in this section. The proposed \secagg scheme, however, works for arbitrary input distributions.}}. This input serves as an abstraction of the locally trained machine learning  models of the clients in  the context of federated learning. To facilitate information-theoretic converse on the relevant \comm and secret key rates, we assume that the inputs of different users are mutually \indep, \ie, 
\begin{align}
\label{eq:input independence}
H(W_k) & = L\, \trm{(in $q$-ary unit)},\; \forall  k\in[K]\notag \\
H(W_{1:K}) &=\sum_{k=1}^KH(W_k).
\end{align} 
To ensure security, each user  also possesses a secret key $Z_k$---referred to as an \emph{\indiv key}---comprising $\lz$ symbols drawn from $\mbb{F}_q$. The keys are assumed to be  \indep  of the inputs, namely,
\be 
\label{eq: indep. between keys and inputs}
H(Z_{1:K}, W_{1:K}) = H(Z_{1:K}) + \sum_{k=1}^KH(W_k).
\ee 
The \indiv keys $Z_1, \cdots,Z_K$ can be arbitrarily correlated and are generated from a \emph{source key} variable $\zsigma$, which consists of $\lzsigma$ symbols, such that
\be 
\label{eq: H(Z1,...,ZK|Zsigma)=0}
H\left(Z_{1:K}|\zsigma\right)=0.
\ee 
The source key $\zsigma$ represents the complete set of independent random symbols required to derive all individual keys. These individual keys are distributed by a trusted third-party entity in an offline fashion (see Remark~\ref{remark: key distn overhead} at the end of this section); specifically, each $Z_k$ is delivered to user $k$ privately before the input aggregation process starts. 

To aggregate the inputs, each user $k\in[K]$ generates a \msg  $X_k$  of length $L_X$ symbols, using its own input and key. The \msg $X_k$ is then  broadcast  to  the remaining $K-1$ users. Specifically, we assume $X_k$ is a deterministic function of $W_k$ and $Z_k$,  so that
\be
\label{eq:H(Xk|Wk,Zk)=0}
H(X_k|W_k,Z_k)=0,\; \forall  k\in[K]
\ee
We further assume that  the $K$ \msgs are transmitted over orthogonal channels (\eg, via TDMA, OFDM, etc.), so that  no interference occurs among them.
Upon  receiving the \msgs from all other users---and with access to its own input and secret key---each user $k$ should be able to recover the sum of all inputs\footnote{In FL, users may want to recover a weighted sum of the inputs, where the weights correspond to the (normalized) dataset sizes of the users. However, these weights can be absorbed into the inputs, allowing the task to be reduced to recovering an unweighted sum,  as given in (\ref{eq:recovery constraint}).}, \ie,
\begin{align}
\label{eq:recovery constraint}
& \mathrm{[Recovery]} \quad H\left(\inputsum \bigg|\{X_i\}_{i\in [K]\bksl \{ k\} }, W_k,Z_k\right  )=0,\notag\\
& \hspace{6.5cm}\forall  k\in[K]
\end{align}
Note that $X_k$ is not included in the  conditioning terms because it is deterministically determined by $W_k$ and $Z_k$.
Moreover, since $W_k$ is available to User $k$, it suffices for the user to recover $\sum_{i \in [K] \setminus \{k\}} X_i$, from which the global sum can be easily computed.

\tbf{Security model.}
{We employ the honest-but-curious security model where each user must not learn any \info about other users' inputs beyond  the aggregate sum (and any information available through collusion), even when colluding with up to $T$ other users.} Specifically, each  user $k$ may collude---\ie, gain access to the  inputs and keys---with any subset of users $\Tc \subset [K]\bksl \{k\}$ with $|\Tc|\le T$. $\Tc$ provides a  general modeling of the security-compromised clients in practical FL systems ({Note that users in $\Tc$ do not collude unless they lie in each other's collusion set}).
\Diff users may potentially collude with \diff subsets of users. For ease of notation, denote 
\be
\label{eq:def C_T}
\Cc_{\Sc} \eqdef \{W_i,Z_i\}_{i\in \Sc}
\ee
as the collection of inputs and keys at a subset $\Sc \subseteq [K]$ of users. Using \muinfo,  the security constraint
can be expressed as
\begin{align}
\label{eq:security constraint}
& \mathrm{[Security]}\quad  \forall \Tc \subset [K]\bksl \{k\}, |\Tc|\le T, \forall k\in[K]: \notag\\
& I\left( \{X_i\}_{i \in [K]\bksl \{k\}}; \{W_i\}_{i \in [K]\bksl \{k\}} \Big|
 \sum_{i=1}^K W_i, 
 W_k, Z_k, \Cc_\Tc   \right)=0,
\end{align}
which implies the statistical \indepce between  the received \msgs by User $k$ and all other users' inputs  when conditioned on User $k$'s own input, secret key and the desired input sum. Note that  the above security constraint needs to be satisfied simultaneously at all users, which implicitly requires the presence of at least $K\ge 3$ users. 
This is because when there are only two users, each user can always recover the other user's input by simply subtracting its own input from the input sum, {in which case no meaningful security can be achieved.  A formal explanation of DSA with $K=2$ users or $T\ge K-2$ colluding users is provided in Section~\ref{subsec:infeasibility proof, converse}.}

\tbf{Performance metrics.}
We study both the \comm and secret key generation aspects of  \dsa. \Msp, the  per-user 
\comm rate $ \rx  $  quantifies the communication efficiency  of  the \agg  process, and is defined as
\be
\label{eq: def comm rate}
R_X \eqdef {L_X}/{L},
\ee 
where $L_X$ denotes the number of symbols contained in each message  $X_k$  for all $k\in[K]$\footnote{Due to symmetry  of the network architecture and \comm protocol, it is assumed that each user's \msg contains the same number of symbols.}. Equivalently, each  user needs to transmit $\rx$  symbols in order 
to recover one symbol of the desired input sum.
The \indiv key rate $R_Z$ quantifies the number of \indep  secret key symbols that each user must possess to ensure security, and is defined as
\be
\label{eq: def indiv key rate}
R_Z \eqdef  {L_Z}/{L},
\ee 
where $L_Z$ represents the number of symbols contained in each $Z_k, \forall k\in[K]$.
The \skr  quantifies the total number of \indep key symbols held collectively by all users, and is defined as 
\be
\label{eq: def source key rate} \rzsigma \eqdef  {\lzsigma}/{L},
\ee 
where $\lzsigma$ represents the number of symbols contained in the source key $\zsigma$. 
A rate triple $(\rx, \rz,\rzsigma)$ is said to be achievable if there exists a secure \agg scheme---namely, the design of the \brdcst \msgs  $X_1,\cdots, X_K$---that \emph{simultaneously} achieves the \comm and key rates $\rx,\rz$, and $\rzsigma$, while satisfying both the recovery and the security constraints (\ref{eq:recovery constraint}), (\ref{eq:security constraint}).
We aim to find the optimal rate region $\Rc^*$, defined as the closure of all achievable rate triples.

\begin{remark}[\Comm \Ovhd of Key Distribution]
\label{remark: key distn overhead}
We assume the presence of a trusted third-party entity (\ie, a key server) responsible for generating and distributing  the individual keys.
Each \indiv key $Z_k$ must be delivered to its corresponding User $k, \forall  k\in[K]$, via a private unicast link, since exposing any individual key to unauthorized users inevitably compromises security. Accordingly, the communication overhead incurred by  key \distn---defined as the total number of symbols sent by the trusted entity (normalized by the input size $L$), is quantified by $  K\rz$.\footnote{{Although the key \distn overhead scales linear with $K$,  it is negligible compared with the \comm savings of the proposed \schm  over the naive baseline presented in Appendix \ref{sec:baseline scheme,appendix}.}} As a result, a smaller key rate  directly leads  to reduced overhead in the key distribution process, which serves as the primary motivation for using $\rz$ as a performance metric DSA.
\end{remark}

\section{Main Result}
\label{sec:main result}
This section presents the main result of the paper, which  characterizes the optimal rate region for the proposed \dsa problem. The result consists of an intuitive \agg scheme and a matching converse proof.
We further highlight the significance of these results by discussing  the \comm and  key generation efficiency of the proposed design.

\begin{theorem}
\label{thm:main result}
{For the \dsa problem with $K\ge  3$ users and at most $ T \le  K- 3$ colluding users, the optimal rate region is given by}
\be 
\label{eq:optimal rate region}
\Rc^* =\left\{ \left(\rx, \rz,\rzsigma\right) \left|  \begin{array}{c}
\rx \ge 1, \\
\rz \ge 1, \\
\rzsigma  \ge K-1
\end{array}
\right.   \right\}.
\ee
\end{theorem}

\begin{IEEEproof}
The proof of \Thm~\ref{thm:main result} consists of an \achvb scheme presented in Section~\ref{sec: ach scheme}, and a tight converse detailing the lower bounds on the   rates of any  \achvb schemes presented in Section~\ref{sec: converse}.
\end{IEEEproof}

We highlight the implications of \Thm~\ref{thm:main result} \af:

\emph{1) {Trivial Cases}}: {When there are only two users, computing the sum $W_1 + W_2$ effectively requires each user to recover the other user's input, leaving nothing to be concealed. Hence, no meaningful security can be provided. Similarly, if a user can collude with $T=K-2$  other users, it will learn the inputs and keys of all users except one. \Aar, the problem reduces to the two-user \agg scenario,  in which no meaningful security can be achieved. Therefore, we consider the case of $K\ge 3$ users and $T\le K-3 $.
}

\emph{2) \Comm and Key Efficiency}: \Thm~\ref{thm:main result} suggests that the optimal rates  of $R_X^*=1, R_Z^*=1$ and $\rzsigmastar=K-1$ are simultaneously \achvb for any $K$. The  optimal \comm and \indiv key rates are  intuitive: for each  input $W_k$ to be aggregated at the remaining users, the message $X_k$ must fully convey the information of $W_k$, implying $L_X\ge H(W_k)$ and thus $\rx\ge 1$. Moreover,  since $W_k$ should be fully protected by $Z_k$ (this a direct consequence of the security constraint (\ref{eq:security constraint})), Shannon's one-time pad (OTP) encryption theorem~\cite{shannon1949communication} states that the key size must be at least as large as the input, \ie, $L_Z\ge H(W_k)$, yielding $ \rz\ge 1$.
In contrast, characterizing the optimal source key rate $\rzsigmastar = K - 1$ is more challenging and relies on the novel converse techniques developed in Section~\ref{sec: converse}.

\emph{3) Connection to Centralized \SecAgg:} \Secagg has been studied in a centralized setting~\cite{zhao2023secure}, where an  \aggtr with no access to inputs and keys computes the sum of inputs from $K$ associated users. 
While DSA  extends this to a \decen setting, major differences exist:
First, in DSA, each user holds both its input and key, which serve as \emph{decryption aids} during aggregation. This is reflected in the conditioning terms $W_k,Z_k$ in the recovery constraint~(\ref{eq:recovery constraint}).
In contrast, the centralized server has no prior knowledge of any inputs or keys, requiring a fundamentally different design.
A natural DSA baseline applies the scheme of~\cite{zhao2023secure} \indeply $K$ times, each time designating one user as the aggregator.\footnote{Refer to Appendix~\ref{sec:baseline scheme,appendix} for a detailed description.} This baseline yields rates $R_X=K-1$, {$R_Z=K$}, and $R_{Z_\Sigma}=K(K-1)$, which are about $K$ times higher than the optimal rates $R_X^*=R_Z^*=1,\rzsigmastar=K-1$. A comparison is given in Table~\ref{tab:comparison of rates}. The substantial reduction in \comm and key rates stems from the optimal key reuse enabled by the proposed scheme, which will be elaborated later.
\begin{table}[t]
\centering 
\caption{Comparison of Rates}
\vspace{-.1cm}
    \begin{tabular}{|c"c|c|c|}
    \hline
    &  $\rx$   & $\rz$  & $\rzsigma$ \\
    \thickhline
       Baseline scheme    &  $K-1$ & {$K$} & $K(K-1)$ \\
     \hline 
      Proposed scheme  & $1$ & $1$  &  $K-1$ \\
     \hline 
    \end{tabular}
    \label{tab:comparison of rates}
    \vspace{-.2cm}
\end{table}
Second, the maximum tolerable collusion threshold differs: centralized  SA supports up to $T = K -2$ colluding users, whereas  DSA tolerates $T = K - 3$. This is because in DSA, each user only recovers the sum of others' inputs (as its own input is already available), while the central server must recover the total sum of all users.

\section{Achievable Scheme}
\label{sec: ach scheme}
Before presenting the general scheme, we first provide a fully worked example to illustrate the key ideas underlying both the proposed design and the associated converse bounds.
\subsection{Motivating Examples}
\label{subsec: example, ach scheme}
\begin{example}
\label{example: ach scheme}
Consider an example with $K=3$ users, as illustrated in Fig.~\ref{fig:model}. Since $T\le K-3=0$,
the example does not tolerate any user collusion.
Suppose each input has $L=1$ bit.
The source key is chosen as 
\be
\label{eq:Zsigma,example}
\zsigma=(N_1,N_2),
\ee 
 where $N_1$ and $N_2$ are two \iid uniform random variables from $\mbb{F}_2$. 
The \indiv  keys of the users are given by
\begin{align}
\label{eq:keys,example,scheme section}
Z_1 = N_1,\;
Z_2  = N_2, \;
Z_3 = -(N_1+N_2).
\end{align}
Hence,  the \indiv and source key rates are $R_Z=1,\rzsigma=2$.
Note that the keys satisfy the zero-sum condition, i.e., $Z_1 + Z_2 + Z_3 = 0$, which is a necessary property to enable input sum recovery. With the above key design, the broadcast messages are chosen as
\begin{align}
\label{eq:msgs,example,scheme section}
X_1 &= W_1+ Z_1 = W_1+N_1, \notag \\
X_2 &= W_2+Z_2=W_2+N_2, \notag \\
X_3 &= W_3 +Z_3=W_3-(N_1+N_2).
\end{align}
Over the binary field $\mbb{F}_2$, the \msgs are just bitwise XORs of the inputs and keys.
Since each \msg contains $L_X=1$ bit, the \comm rate is equal to $R_X=1$.

\tbf{Input sum recovery.} With the above key and \msg design, we show  that each user  can correctly recover the input sum $W_1+W_2+W_3$. Specifically, let us consider User 1. Since User 1 possesses $W_1$, it only needs to recover  $W_2+W_3$. \Ip, it 
computes the sum of the received \msgs $X_2$ and $X_3$ along with its own key $Z_1$ to obtain 
\begin{align}
& X_2+X_3+Z_1   \overset{(\ref{eq:msgs,example,scheme section})}{=}  (W_2+Z_2)+ (W_3+Z_3) +Z_1\notag\\
&\qquad =W_2+W_3+ Z_1+Z_2+Z_3  \overset{(\ref{eq:keys,example,scheme section})}{=}  W_2+W_3,
\end{align}
where the last step is due to the zero-sum property of the  \indiv keys (see (\ref{eq:keys,example,scheme section})).
Similarly, the two remaining users can also recover the desired \insum \af:
\begin{align}
\trm{User  2:}\quad X_1+X_3+Z_2 & = W_1 + W_3,  \notag\\
\trm{User  3:}\quad X_1+X_2+Z_3 & = W_1 + W_2.
\end{align}

\tbf{Proof of security.} We now show that  the proposed scheme satisfies the security constraint (\ref{eq:security constraint}). \Wlog, let us consider User 1. An intuitive proof is provided \af: 
Suppose User 1 recovers the desired input sum
$W_2+W_3$ by linearly combining $X_2,X_3$ and $Z_1$ with coefficients $\ell_1, \ell_2$ and $\ell_3$, \resp. That is,
\begin{align}
&  \ell_1X_2+\ell_2X_3 + \ell_3Z_1\notag\\
& =  {\ell_1}W_2+ {\ell_2}W_3 +  (\ell_3-\ell_2)N_1 +  (\ell_1-\ell_2)N_2.
\end{align}
Because $N_1$ and $N_2$ are \indep random variables, to cancel out  the key component $(\ell_3-\ell_2)N_1 +  (\ell_1-\ell_2)N_2$, the coefficients of $N_1$ and $N_2$ must be zero, \ie,
$ \ell_3-\ell_2=\ell_1-\ell_2=0$, suggesting $\ell_1=\ell_2=\ell_3$.
Therefore, any key-canceling linear combination\footnote{A key-canceling linear combination ensures that all key variables cancel out completely, leaving only the input component intact.} $\bm{\ell}\eqdef (\ell_1, \ell_2, \ell_3)$ must take the form $\bm{\ell}=(\ell, \ell, \ell), \ell \in \{0,1\}$. As a  result,
\be
\label{eq:l*X=l(W2+W3),example}
\bm{\ell}(X_2,X_3,Z_1)^{  T}=\ell(W_2+W_3),
\ee 
which is a function of the desired input sum $W_2+W_3$. Equation (\ref{eq:l*X=l(W2+W3),example})  implies that any key-canceling linear combination of the received \msgs and User 1's key reveals no additional information about  $(W_2, W_3)$ beyond their sum $W_2+W_3$, thereby ensuring security.

A rigorous proof through mutual information is presented below. By (\ref{eq:security constraint}), we have
\begin{align}
\label{eq:proof of security, example}
& I\left(X_2,X_3;W_2,W_3|W_2+W_3, Z_1\right)\notag\\
&  =H(X_2,X_3|W_2+W_3, Z_1) - H(X_2,X_3|W_2,W_3, Z_1).
\end{align}
We now compute the two terms  in (\ref{eq:proof of security, example}) \resp.  First,  the second term is equal to
\begin{subequations}
\label{eq:term 2, proof of security, example}
\begin{align}
& H(X_2,X_3|W_2,W_3, Z_1)\notag \\
& = H\left(W_2+N_2,W_3-(N_1+N_2) |W_2,W_3, N_1\right)  \\
& = H\left(N_2,N_1+N_2 |W_2,W_3, N_1\right)  \\
& \overset{(\ref{eq: indep. between keys and inputs})}{=} H(N_2,N_1+N_2 |N_1) \label{eq:step0,term 2, proof of security, example} \\
& =  H(N_2|N_1) \\
& \overset{(\ref{eq:Zsigma,example})}{=} H(N_2)=1, \label{eq:step1,term 2, proof of security, example}
\end{align}
\end{subequations}
where (\ref{eq:step0,term 2, proof of security, example}) is due to the \indepce of the keys and inputs, and (\ref{eq:step1,term 2, proof of security, example}) is due to the \indepce of the source key symbols.
Second, the first term in (\ref{eq:proof of security, example}) is bounded by  
\begin{subequations}
\label{eq:term 1, proof of security, example}
\begin{align}
& H(X_2,X_3|W_2+W_3, Z_1) \notag  \\
&  = H(W_2+N_2,W_3-(N_1+N_2) |W_2+W_3, N_1)\\
&  = H(W_2+N_2,W_3-N_2 |W_2+W_3, N_1)\\
& \overset{(\ref{eq: indep. between keys and inputs}),(\ref{eq:Zsigma,example})}{=} H(W_2+N_2,W_3-N_2 |W_2+W_3)\label{eq:step 0, term 1}\\
&=H(W_2+N_2,W_3-N_2,W_2+W_3)-H(W_2+W_3)\\
& \overset{(\ref{eq:input independence})}{=}   H(W_2+N_2,W_3-N_2)-1\label{eq:step 1, term 1}\\
&\le H(W_2+N_2) + H(W_3-N_2)-1\le 1,
\end{align}
\end{subequations}
where (\ref{eq:step 0, term 1}) is due to the \indepce of the keys and inputs,  and also the \indepce between $N_1$ and $N_2$.
(\ref{eq:step 1, term 1}) is due to the uniformity of the inputs.
The last step holds because $H(W_2+N_2)\le 1, H(W_3-N_2)\le  1  $, as each of $W_2 +N_2$ and $W_3-N_2$ contains one bit and uniform \distn maximizes the entropy. 

Plugging (\ref{eq:term 1, proof of security, example}) and (\ref{eq:term 2, proof of security, example}) into (\ref{eq:proof of security, example}), we have
\be 
I(X_2,X_3;W_2,W_3|W_2+W_3, Z_1)\le0.\notag
\ee 
Since \muinfo is non-negative, we conclude that $I(X_2,X_3;W_2,W_3|W_2+W_3, Z_1)=0$, proving the security for User 1. 
The proofs for other  users follow similarly.

\tbf{Converse}. 
The above  scheme simultaneously achieves the rates $\rx=1,\rz=1$, and $\rzsigma=2$ . We now show that the following lower bounds
\be
\label{eq:lower bounds,converse,example}
\rx \ge 1,\;  \rz \ge 1, \;\rzsigma  \ge 2
\ee 
must be satisfied for any \secagg scheme. These bounds are established  through entropic arguments based on the  \itic formulations of the recovery constraint  (\ref{eq:recovery constraint}), the security constraint (\ref{eq:security constraint}), the \indepce between inputs and keys (see (\ref{eq: indep. between keys and inputs})), and the uniformity and mutual \indepce of the inputs (see (\ref{eq:input independence})).

We begin with the following intermediate result, which states that the entropy of $X_1$, when conditioned on the inputs and keys of users 2 and 3, must be at least as large as $H(W_1)$:
\be
\label{eq:result1,H(X1|(Wi,Zi)i=2,3)>=L,example}
H\left(X_1|W_2,W_3, Z_2,Z_3\right) \ge H(W_1)=L.
\ee 
This result follows from an intuitive cut-set argument: because $W_1$ resides only at User 1, it must pass through $X_1$ in order for
User 2 (or User 3) to recover $W_1+W_3$ (or $W_2+W_3$). Therefore, the entropy of $X_1$ must be as least as large as that of $W_1$.
More formally, we have
\begin{subequations}
\label{eq:proof of lemma, example}
\begin{align}
& H\left(X_1| W_2,W_3, Z_2,Z_3   \right) \notag \\ 
 & \ge  I\left(X_1; W_1+W_3|\{W_i,Z_i\}_{i \in\{2,3\} }\right)\\
& =  H\left(W_1+W_3|\{W_i,Z_i\}_{i \in\{2,3\} }\right)\notag\\
& \quad - H\left(W_1+W_3|\{W_i,Z_i\}_{i \in\{2,3\}},X_1\right) \\
& =  H\left(W_1\right) - H\left(W_1+W_3|\{W_i,Z_i\}_{i \in\{2,3\}},X_1,X_3\right)
\label{eq:step 0, proof of lemma, example}\\
& \ge  H\left(W_1\right) -   
\underbrace{ H\left(W_1+W_3|W_2,Z_2,X_1,X_3\right)}_{\overset{(\ref{eq:recovery constraint})}{=} 0  }\\
& =   H(W_1)=L,\label{eq:step 1, proof of lemma, example}
\end{align}
\end{subequations}
where (\ref{eq:step 0, proof of lemma, example})  is  due to the independence of the inputs from the keys, and also because $X_3$ is a deterministic function of $W_3$ and $Z_3$.  Moreover, (\ref{eq:step 1, proof of lemma, example}) is  due to the recovery constraint of User 2 (see (\ref{eq:recovery constraint})). Similarly, we can also obtain
\be
\label{eq:result1,H(X2|(Wi,Zi)i=1,3)>=L,example}
H\left(X_2|W_1,W_3, Z_1,Z_3\right) \ge H(W_2)=L.
\ee 

\tit{1) Proof  of $R_X\ge 1$.}
Equipped with  (\ref{eq:result1,H(X1|(Wi,Zi)i=2,3)>=L,example}), $R_X\ge 1$ follows immediately: 
\begin{align}
 H(X_1) & \ge H(X_1|W_2,W_3, Z_2,Z_3) 
 \overset{(\ref{eq:result1,H(X1|(Wi,Zi)i=2,3)>=L,example})}{\ge} L,\notag\\
& \Rightarrow \rx \eqdef {L_X}/{L}  \ge {H(X_1)}/{L}\ge 1, 
\end{align}
where the first inequality follows from the fact that conditioning cannot increase entropy. 
Note that  $L_X$ denotes the number symbols contained in $X_1$, which is always lower bounded by the entropy of $X_1$. Intuitively, $\rx \ge 1$ holds because each \msg $X_k$ must fully carry the \info of the corresponding input $W_k$ so that $L_X\ge L,\forall k\in[3]$.

To derive lower bounds on the \indiv and source key rates, we first introduce the following result,  which asserts that the broadcast message $X_1$  of  User 1  must be \indep  of its own input $W_1$, when conditioned on the input and key of any another user:
\be
\label{eq:I(X1;W1|Wk,Zk)=0,indetermediate result,example}
I\left( X_1; W_1 | W_k, Z_k\right)=0,\; \forall k \in\{2,3\}
\ee 
This is a direct consequence of the security constraint (\ref{eq:security constraint}) imposed at  both User 2 and 3. Specifically, since $X_1$ is observed by both users, $W_1$ must be fully protected (by $Z_1$)  to ensure that $X_1$ appears  \indep of $W_1$ from the perspectives of Users 2 and 3.
\Wlog, let us consider $k=2$:
\begin{subequations}
\label{eq:proof,I(X1;W1|Wk,Zk)=0,indetermediate result,example}
\begin{align}
& I\left( X_1; W_1 | W_2, Z_2\right)\notag \\
& \le I\left( X_1,W_1+W_2+W_3; W_1 | W_2, Z_2\right)\\
&  = I\left( W_1+W_2+W_3; W_1| W_2, Z_2\right)\notag\\
& \quad  +  I\left( X_1; W_1|W_1+W_2+W_3, W_2, Z_2\right)\\
& \le  I\left( W_1+W_3; W_1| W_2, Z_2\right)\notag\\
& \quad  + \underbrace{I\left( X_1,X_3 ; W_1,W_3|W_1+W_2+W_3, W_2, Z_2\right)}_{\ovst{(\ref{eq:security constraint})}{=}0  }\label{eq:step0,proof,I(X1;W1|Wk,Zk)=0,indetermediate result,example}\\
& =  I\left( W_1+W_3; W_1| W_2, Z_2\right)\\
& \ovst{(\ref{eq: indep. between keys and inputs})}{=}  I\left( W_1+W_3; W_1| W_2\right)\label{eq:step1,proof,I(X1;W1|Wk,Zk)=0,indetermediate result,example}\\
& \ovst{(\ref{eq:input independence})}{=}  I\left( W_1+W_3; W_1\right)\label{eq:step2,proof,I(X1;W1|Wk,Zk)=0,indetermediate result,example}\\
& = H(W_1+W_3)-H(W_3)\\
& =L-L=0,
\end{align}
\end{subequations}
where (\ref{eq:step0,proof,I(X1;W1|Wk,Zk)=0,indetermediate result,example}) is due to the security constraint  of User 2.  (\ref{eq:step1,proof,I(X1;W1|Wk,Zk)=0,indetermediate result,example}) and (\ref{eq:step2,proof,I(X1;W1|Wk,Zk)=0,indetermediate result,example}) are  due to  the \indepce  of the inputs and keys (see (\ref{eq: indep. between keys and inputs})), and the \indepce among the inputs (see (\ref{eq:input independence})), \resp. Since \muinfo cannot be negative, we conclude that $I\left( X_1; W_1 | W_2, Z_2\right)=0$. Similarly, we can show  $I\left( X_1; W_1 | W_3, Z_3\right)=0$ due to the security at User 3. Based on (\ref{eq:proof,I(X1;W1|Wk,Zk)=0,indetermediate result,example}), we prove the lower bounds on the key rates  \af:

\tit{2) Proof of $R_Z\ge 1$.} Consider $Z_1$. We have
\begin{subequations}
\label{eq:proof Rz>=1,example}
\begin{align}
L_Z &  \ge H(Z_1)\label{eq:step-1,proof Rz>=1,example}\\
& \ge H\left(Z_1 | W_1,W_2,Z_2 \right)\\
& \ge I\left(Z_1;X_1 | W_1,W_2,Z_2 \right)\\
& = H\left(X_1| W_1,W_2,Z_2 \right)
- \underbrace{H\left(X_1| W_1,W_2,Z_1, Z_2 \right)}_{\ovst{(\ref{eq:H(Xk|Wk,Zk)=0})}{=}0}  \label{eq:step0,proof Rz>=1,example}\\
& = H\left(X_1|W_2,Z_2 \right) - \underbrace{I\left(X_1;W_1|W_2,Z_2 \right)}_{\ovst{(\ref{eq:I(X1;W1|Wk,Zk)=0,indetermediate result,example})}{=}0} \label{eq:step1,proof Rz>=1,example}\\
& \ge H\left(X_1|W_2,Z_2, W_3,Z_3 \right)\label{eq:step11,proof Rz>=1,example}\\
&  \ovst{(\ref{eq:result1,H(X1|(Wi,Zi)i=2,3)>=L,example})}{\ge} L\Rightarrow \rz \eqdef {L_Z}/{L}\ge 1,  \label{eq:step2,proof Rz>=1,example}
\end{align}
\end{subequations}
where (\ref{eq:step0,proof Rz>=1,example})  is because $X_1$ is a deterministic function of $W_1$ and $Z_1$ so that $H(X_1 |W_1,Z_1)=0$ (see (\ref{eq:H(Xk|Wk,Zk)=0})). In (\ref{eq:step1,proof Rz>=1,example}),  we applied the result from (\ref{eq:I(X1;W1|Wk,Zk)=0,indetermediate result,example}). In (\ref{eq:step2,proof Rz>=1,example}), (\ref{eq:result1,H(X1|(Wi,Zi)i=2,3)>=L,example}) was applied. 
Intuitively, the bound $\rz \ge 1$ follows from the requirement that each input $W_k$ must be protected by its corresponding key $Z_k$. According to Shannon's one-time pad (OTP) encryption theorem~\cite{shannon1949communication}, the key size must be at least as large as the message size to ensure perfect secrecy, as demonstrated in (\ref{eq:step-1,proof Rz>=1,example})-(\ref{eq:step11,proof Rz>=1,example}). Since each message must contain at least $L$ symbols, the key must also be no shorter than $L$.

\tit{3) Proof of $\rzsigma\ge 2$.}
To prove this, we first establish the following bounds:
\begin{align}
H(X_1, X_2|W_3, Z_3)  &\ge  2L,
\label{eq:H(X1,X2|W3,Z3)>=2L,example} \\
I(X_1, X_2;W_1, W_2|W_3, Z_3) & =L, \label{eq:I(X_1,X_2;W_1,W_2|W_3,Z_3)=L}\\
H(Z_1,Z_2|Z_3) & \ge  L.\label{eq:H(Z1,Z2|Z3)>=L,example}
\end{align}
(\ref{eq:H(X1,X2|W3,Z3)>=2L,example}) shows that the joint entropy of the \msgs $X_1,X_2$ conditioned on the input and key of User 3 is  at least $2L$. Intuitively, this is because the inputs $W_1$ and $W_2 $ must  be  incorporated in $X_1$ and $X_2$ \resp, so that the two \msgs' joint entropy must be at least $H(W_1,W_2)=2L$. Rigorously, we have
\begin{subequations}
\label{eq:proof H(X1,X2|W3,Z3)>=2L,example}
\begin{align}
& H(X_1, X_2|W_3, Z_3) \notag\\
& = H(X_1|W_3, Z_3) + H(X_2|W_3, Z_3, X_1) \\
&  \ge H(X_1|W_2, Z_2,W_3, Z_3) + H(X_2|W_3, Z_3, W_1, Z_1, X_1) \\
&  \overset{(\ref{eq:H(Xk|Wk,Zk)=0})}{=} H(X_1|W_2, Z_2,W_3, Z_3) + H(X_2|W_3, Z_3, W_1, Z_1) \\
& \ovst{(\ref{eq:result1,H(X1|(Wi,Zi)i=2,3)>=L,example})(\ref{eq:result1,H(X2|(Wi,Zi)i=1,3)>=L,example})}{\ge }2L,\label{eq:step0,proof H(X1,X2|W3,Z3)>=2L,example}
\end{align}
\end{subequations}
where  (\ref{eq:result1,H(X1|(Wi,Zi)i=2,3)>=L,example}) and (\ref{eq:result1,H(X2|(Wi,Zi)i=1,3)>=L,example}) are applied in the last step.

Moreover,  (\ref{eq:I(X_1,X_2;W_1,W_2|W_3,Z_3)=L}) shows that the \muinfo between $(X_1,X_2)$ and $(W_1,W_2)$---when conditioned on $W_3$ and $Z_3$, is at least $L$. This is because from User 3's perspective, the only thing that can be inferred from $X_1$ and $X_2$ is $W_1+W_2$, which consists of $L$ symbols due to the uniformity of the inputs. Specifically, we have
\begin{subequations}
\label{eq:proof,I(X_1,X_2;W_1,W_2|W_3,Z_3)=L}
\begin{align}
&I(X_1,X_2;W_1,W_2|W_3,Z_3) \notag\\
& = I(X_1,X_2;W_1,W_2,W_1+W_2|W_3,Z_3)\\
& =I(X_1,X_2;W_1+W_2|W_3,Z_3)\notag\\
& \quad +  \underbrace{
I(X_1,X_2;W_1,W_2|W_1+W_2,W_3,Z_3)}_{\ovst{(\ref{eq:security constraint})}{=}0  }\label{eq:step0,proof,I(X_1,X_2;W_1,W_2|W_3,Z_3)=L}\\
& =H(W_1+W_2|W_3,Z_3) -  H(W_1+W_2|W_3,Z_3,X_1, X_2)\\
& \ovst{(\ref{eq: indep. between keys and inputs}),(\ref{eq:recovery constraint})}{=}    H(W_1+W_2|W_3)\label{eq:step1,proof,I(X_1,X_2;W_1,W_2|W_3,Z_3)=L}\\
&  \ovst{(\ref{eq:input independence})}{=} H(W_1+W_2)=L,
\end{align}
\end{subequations}
where (\ref{eq:step0,proof,I(X_1,X_2;W_1,W_2|W_3,Z_3)=L}) and (\ref{eq:step1,proof,I(X_1,X_2;W_1,W_2|W_3,Z_3)=L}) are \resp  due to the security and recovery constraints of User 3 (see (\ref{eq:security constraint}), (\ref{eq:recovery constraint})). The last two steps  are due to the \indepce of the keys and inputs, and also the uniformity of the inputs.

Based on (\ref{eq:H(X1,X2|W3,Z3)>=2L,example}) and (\ref{eq:I(X_1,X_2;W_1,W_2|W_3,Z_3)=L}), 
(\ref{eq:H(Z1,Z2|Z3)>=L,example}) establishes a lower bound $L$ 
on the joint entropy of $Z_1$ and $Z_2$ conditioned on $Z_3$, 
which is proved \af:
\begin{subequations}
\label{eq:proof,H(Z1,Z2|Z3)>=L,example}
\begin{align}
& H(Z_1,Z_2|Z_3) \ovst{(\ref{eq: indep. between keys and inputs})}{=}   H(Z_1,Z_2|W_1, W_2 , W_3,Z_3)\label{eq:step0,proof,H(Z1,Z2|Z3)>=L,example}\\
&\ge  I(Z_1,Z_2; X_1,X_2|W_1, W_2,W_3,Z_3)\\
& = H(X_1,X_2|W_1, W_2,W_3,Z_3)  \notag\\
& \quad  - \underbrace{H(X_1,X_2|W_1, W_2,W_3,Z_3, Z_1,Z_2)}_{\ovst{(\ref{eq:H(Xk|Wk,Zk)=0})}{=}0}\label{eq:step1,proof,H(Z1,Z2|Z3)>=L,example}\\
& =H(X_1, X_2|W_3, Z_3 ) -  I(X_1, X_2;W_1,W_2|W_3, Z_3) \\
&  \ovst{(\ref{eq:H(X1,X2|W3,Z3)>=2L,example}),(\ref{eq:I(X_1,X_2;W_1,W_2|W_3,Z_3)=L})}{ \ge } L ,
\end{align}
\end{subequations}
where (\ref{eq:step0,proof,H(Z1,Z2|Z3)>=L,example}) is due to the \indepce  of the inputs and keys. In the last step, we applied the two previously obtained bounds (\ref{eq:H(X1,X2|W3,Z3)>=2L,example}) and (\ref{eq:I(X_1,X_2;W_1,W_2|W_3,Z_3)=L}).  

With the above intermediate results (\ref{eq:H(X1,X2|W3,Z3)>=2L,example})-(\ref{eq:H(Z1,Z2|Z3)>=L,example}), we are now ready to prove $\rzsigma \ge 2$:
\begin{subequations}
\label{eq:proof of Rzsigma>=2}
\begin{align}
\lzsigma \ge H\left( \zsigma  \right) &  \ovst{(
 \ref{eq: H(Z1,...,ZK|Zsigma)=0}
 )}{=}H\left( \zsigma  \right) + H\left(Z_1, Z_2,Z_3| \zsigma  \right)\label{eq:step0,proof of Rzsigma>=2}\\
 & = H\left(Z_1, Z_2,Z_3,\zsigma  \right)\\
  & \ge H\left(Z_1, Z_2,Z_3  \right)\\
  & \ge H\left(Z_3  \right) + H\left(Z_1, Z_2|Z_3  \right) \\
 & \ovst{(\ref{eq:proof Rz>=1,example}),(\ref{eq:H(Z1,Z2|Z3)>=L,example})}{\ge } 2L,\label{eq:step1,proof of Rzsigma>=2}\\
& \Rightarrow \rzsigma \eqdef {\lzsigma}/{L}\ge 2,
\end{align}
\end{subequations}
where (\ref{eq:step0,proof of Rzsigma>=2})  is because the \indiv keys are generated from the source key.  In (\ref{eq:step1,proof of Rzsigma>=2}), we used the \indiv key bound (\ref{eq:proof Rz>=1,example}) and (\ref{eq:H(Z1,Z2|Z3)>=L,example}). \Aar, we have proved $\rzsigma  \ge 2$. 

Since  the \achvb rates of the proposed \agg scheme in Example~\ref{example: ach scheme} match the derived lower bounds, the optimal rate region  is given by $\Rc^* =\{(\rx,\rz,\rzsigma): \rx \ge 1, \rz\ge 1, \rzsigma \ge 2\}$.
\hfill  $\lozenge$
\end{example}

\begin{remark}[$\Rc^*$ as a Box Region]
The rates defined in (\ref{eq: def comm rate}), (\ref{eq: def indiv key rate}) and (\ref{eq: def source key rate}) are based on the number of symbols $\lx, \lz$, and $\lzsigma$ in each message, \indiv key, and source key \resp. These quantities can be made arbitrarily large by padding zeros to the messages and keys. However, such padding does not affect  their entropy. As a result, the optimal rate region $\Rc^*$---as defined via (\ref{eq: def comm rate}), (\ref{eq: def indiv key rate}) and (\ref{eq: def source key rate})---can be conveniently expressed as a box region. This avoids complications that may arise from potential correlations among the rates if they are instead defined  in terms of entropy, \ie, $R_X^\prime \eqdef H(X_k)/L, R_Z^\prime \eqdef H(Z_k)/L $, and $R_{Z_\Sigma}^\prime \eqdef H(\zsigma)/L$. \Msp, by Shannon's one-time-pad encryption theorem~\cite{shannon1949communication}, it must hold that $H(Z_k)\ge  H(X_k),\forall k\in[K]$,  implying that $ R_Z^\prime \ge  R_X^\prime$. Under this entropic definition, $\Rc^*$ is no longer a box region.
\end{remark}

\subsection{General Scheme}
\label{subsec: general scheme}
The general scheme is described \af.
Suppose each input has $L=1$ uniform symbol from some finite field $\mbb{F}_q$. The source key is given by 
\be
\label{eq: source key, gen scheme}
\zsigma =\left(N_1, N_2,\cdots, N_{K-1}\right),
\ee 
where $N_1, \cdots, N_{K-1}$ are $K-1$ \iid uniform random variables from the same  field $\mbb{F}_q$. Therefore, the source key rate is equal to $\rzsigma=H(N_1, \cdots, N_{K-1})/L=K-1$. The \indiv keys are chosen as 
\begin{align}
\label{eq: key design, gen scheme}
Z_k & = N_k, \; k \in[K-1] \notag  \\
Z_K & = -(N_1+\cdots+ N_{K-1}).
\end{align}
This key design has {\tit{two}} properties: first, the \indiv keys have a zero sum, \ie, $Z_1+\cdots+Z_{K}=0$; second, any subset of no more than $K-1$ keys are mutually \indep. The first property is necessary for  the input sum recovery {at each user while the second property is necessary for security. \Ip, the first property ensures the cancellation of the keys when the received \msgs are aggregated, therefore exposing the desired input sum. The second property ensures that no unintended input \info is exposed to any user if the user manipulates with the received \msgs, therefore guaranteeing security.}
With the above key design, User $k\in [K]$ broadcasts
\be
\label{eq:msg design, gen scheme}
X_k = W_k + Z_k
\ee 
to all other users.

\tbf{Input Sum Recovery}.
To recover the input  sum, each user adds up all received \msgs and plugs in its own input and secret key to decode  the desired input sum.  \Ip,  User  $k$ performs 
\begin{align}
\label{eq:input sum recovery,gen scheme}
 &\left( X_1 + \cdots+X_{k-1}+ X_{k+1}+\cdots + X_K \right) + W_k +Z_k    \notag \\
 & \;  = \sum_{k=1}^KW_k + \sum_{k=1}^K Z_k \overset{(\ref{eq: key design, gen scheme})}{=} \sum_{k=1}^KW_k
\end{align}
to recover the  sum of inputs of all users.

\tbf{Performance}.  The \achved rates are $\rx=1, \rz=1$ and $\rzsigma=K-1$.

\begin{remark}[Dual Role of \Indiv Keys]
Note that the key design in (\ref{eq: key design, gen scheme}) has an Maximum-Distance-Separable (MDS) property that any subset of $K-1$ \indiv keys are mutually \indep, while the sum of all $K$ keys equals zero.
\Aar, User $k$ cannot decode the desired input sum $\sum_{i\in [K]\bksl \{k\}  }W_i$ from $ \sum_{i\in [K]\bksl \{k\}  }X_i  $ without the help of its own key  $Z_k$, as shown in (\ref{eq:input sum recovery,gen scheme}). This highlights the \emph{dual} role of the \indiv keys in \dsa: on the one hand, $Z_k$ encrypts $W_k$ through $X_k=W_k+Z_k$; on the other hand, $Z_k$ acts as a decryption aid (or `antidote') to recover the desired input sum embedded in the  received \msgs $\{X_i\}_{i\in[K]\bksl\{k\}}$, as shown in (\ref{eq:input sum recovery,gen scheme}).
A fundamental insight of this work is that \emph{the same key can simultaneously fulfill both roles}, without requiring additional key symbols compared to centralized \secagg~\cite{zhao2023secure}. 
\end{remark}

\tbf{Proof of Security}.
With  the proposed secret key and \msg design, we prove that the security constraint in (\ref{eq:security constraint}) can be simultaneously satisfied for all users. Specifically,  consider User $k$ and any  colluding user set $\Tc \subset [K]\bksl\{k\}$ where $|\Tc|\le K-3$. \Feon, let us denote $N_K \eqdef -(N_1+\cdots+N_{K-1})$ so that  $Z_k=N_k, \forall k\in[K]$, and $\overline{\Tc}\eqdef ([K]\bksl \{k\})\bksl \Tc$. 
We have
\begin{subequations}
\label{eq:proof of security, general scheme}
\begin{align}
& I\left( \{X_i\}_{i \in [K]\bksl \{k\}}; \{W_i\}_{i \in [K]\bksl \{k\}} \bigg|
 \sum_{i=1}^K W_i, 
 W_k, Z_k, \Cc_\Tc   \right)\notag \\
 &\overset{(\ref{eq: key design, gen scheme})}{=}    I\left( \{W_i+N_i\}_{i \in \overline{\Tc}   }; \{W_i\}_{i\in \overline{\Tc}} \bigg|
 \sum_{i \in  \overline{\Tc}  } W_i, 
 \Cc_{\Tc \cup  \{k\}   }   \right)\label{eq:step0,proof of security, general scheme} \\
 & 
= H\left( \{W_i+N_i\}_{i \in \overline{\Tc}   } \bigg|
 \sum_{i \in  \overline{\Tc}  } W_i, 
 \Cc_{\Tc \cup  \{k\}   }   \right)\notag\\
& \;\;- H\left( \{W_i+N_i\}_{i \in \overline{\Tc}   } \bigg|
 \sum_{i \in  \overline{\Tc}  } W_i, 
 \Cc_{\Tc \cup  \{k\}   }, \{W_i\}_{i\in \overline{\Tc}}   \right)\\
& 
\overset{(\ref{eq: key design, gen scheme})}{=}  H\left( \{W_i+N_i\}_{i \in \overline{\Tc}   } \bigg|
 \sum_{i \in  \overline{\Tc}  } W_i, \sum_{i \in \overline{\Tc}}N_i,  \Cc_{\Tc \cup  \{k\}   }   \right)\notag\\
& \quad 
- H\left( \{N_i\}_{i \in \overline{\Tc}   } \bigg|
 \sum_{i \in  \overline{\Tc}  } W_i, 
 \Cc_{\Tc \cup  \{k\}   }, \{W_i\}_{i\in \overline{\Tc}}   \right)\label{eq:step1,proof of security, general scheme}\\
& \overset{(\ref{eq: indep. between keys and inputs})}{\le }  
H\left( \{W_i+N_i\}_{i \in \overline{\Tc}   } \bigg|
 \sum_{i \in  \overline{\Tc}  } (W_i+N_i)  \right)\notag\\
& \quad 
- H\left( \{N_i\}_{i \in \overline{\Tc}   } \big|
 \{N_i\}_{i\in \Tc \cup  \{k\}   }   \right)\label{eq:step2,proof of security, general scheme}\\
& =  
H\left( \{W_i+N_i\}_{i \in \overline{\Tc}   } \right) 
-
H\left(  \sum_{i \in  \overline{\Tc}  } (W_i+N_i)\right)
\notag\\
& \quad 
- 
H\left( \{N_i\}_{i \in [K]   }   \right)
+H\left(
 \{N_i\}_{i\in \Tc \cup  \{k\}   }   \right)
 \label{eq:step3,proof of security, general scheme}\\
 & 
 \overset{(\ref{eq:input independence}), (\ref{eq: indep. between keys and inputs}),(\ref{eq: key design, gen scheme})}{=}
|\overline{\Tc}|L-L-(K-1)L+|\Tc \cup \{k\}|L \label{eq:step4,proof of security, general scheme}\\
&=
\left(|\overline{\Tc}|+|\Tc \cup \{k\}|-K     \right)L\\
& =0,
\end{align}
\end{subequations}
where in (\ref{eq:step0,proof of security, general scheme}), we substitute the message and key constructions from  (\ref{eq:msg design, gen scheme}) and (\ref{eq: key design, gen scheme}), \resp, and exclude the conditioning inputs and keys $ \Cc_{\Tc \cup  \{k\}   }= \{W_i,N_i\}_{i\in \Tc\cup  \{k\} }$ from the relevant terms. (\ref{eq:step1,proof of security, general scheme}) is due to the zero-sum property of the keys, \ie,
$\sum_{i \in \overline{\Tc}}N_i= - \sum_{i\in \Tc \cup  \{k\}} N_i  $, which can be derived from $\Cc_{\Tc\cup \{k\}}$. (\ref{eq:step2,proof of security, general scheme}) is due to the \indepce of the inputs and keys (see (\ref{eq: indep. between keys and inputs})). 
(\ref{eq:step3,proof of security, general scheme})
is because 
$H\big(  \sum_{i \in  \overline{\Tc}  } (W_i+N_i)| \{W_i+N_i\}_{i \in \overline{\Tc}   }\big)=0$, and also  the fact that  $ \Tc \cup \overline{\Tc}\cup  \{k\}=[K]$. (\ref{eq:step4,proof of security, general scheme})  follows from the uniformity of both the inputs and keys, as well as their mutual independence. Since \muinfo is non-negative, we have $I\big( \{X_i\}_{i \in [K]\bksl \{k\}}; \{W_i\}_{i \in [K]\bksl \{k\}}|
 \sum_{i=1}^K W_i, 
W_k, Z_k, \Cc_\Tc \big)=0$. Since this condition holds simultaneously for all users, security is proved.

\section{Converse: Lower Bounds on \Comm  and Key Rates}
\label{sec: converse}
In this section, we provide a comprehensive entropic  proof to establish the lower  bounds
\be
\rx  \ge 1, \rz \ge 1, \rzsigma \ge K-1
\ee 
on the \comm and key rates. Since  these converse bounds match the rates  \achved by the proposed scheme described in Section~\ref{subsec: general scheme}, the optimal rate region $\Rc^*$ can be  determined, as shown  in (\ref{eq:optimal rate region}).

\subsection{{Trivial Cases: $K=2$ or $T\ge K-2$}}
\label{subsec:infeasibility proof, converse}
{Note that no meaningful security can be achieved  when $K=2$ or $T\ge K-2$.
As the  case of $2$ users has already been explained in the first implication of \Thm \ref{thm:main result}, we consider here the case when the number of colluding users exceeds $K-3$. For example,
let us consider User $k$ and $|\Tc|=K-2$ colluding users. In this case, the colluding user set can be written as $\Tc=[K]\bksl \{k,k^\prime\}$ where $k^\prime  \in [K]\bksl\{k\}$. To recover the input sum $\sum_{i\in [K]}W_i$ at User $k$, User $k^\prime$'s input $W_{k^\prime}$ will be exposed to User $k$ as it already knows every other summand $\{W_i\}_{i\in  [K]\bksl \{k^\prime\} }$ through collusion. That is, with $T=K-2$ colluding users, the problem reduces to the two-user  \secagg case where no meaningful security can be achieved.}
\Iwf,  we assume that $K\ge 3$ and $T\le K-3$ so that \dsa remains nontrivial.

\subsection{Key Lemmas and Corollaries}
\label{subsec:lemmas&corollaries}
In this section, we present several lemmas and corollaries that play a central role in the converse proof. These results precisely characterize the simultaneous recovery and security constraints imposed on all users, forming the foundation for deriving tight converse bounds on the communication and key rates. Each lemma is stated individually, accompanied by a detailed proof and an intuitive interpretation.

{The first lemma states that each message $X_k$ contains at least $L$ independent symbols, even when the inputs and keys of all other users are known.}

\begin{lemma}
\label{lemma: H(Xk|(Wi,Zi) all other i)}
\emph{For any $k\in [K]$,  it holds that
\be 
\label{eq: H(Xk|(Wi,Zi) all other i), lemma}
H\left(X_k| \{W_i, Z_i\}_{i\in [K]\bksl \{k\}  } \right) \ge H(W_k)=L.   
\ee 
}
\end{lemma}
\begin{IEEEproof}
Let $k^\prime\in [K]\bksl \{k\}$ be any user other than User $k$. Consider  the input \agg at User $k^\prime$, as illustrated in Fig.~\ref{fig:fig1}.
\begin{figure}[ht]
    \centering
    \includegraphics[width=0.28\textwidth]{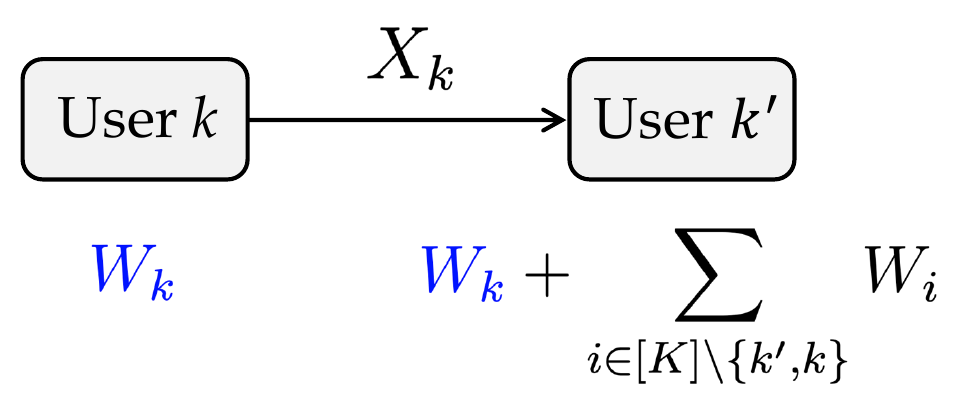}
    \vspace{-.3cm}
    \caption{\Agg of User $k$'s input $W_k$ at User $k^\prime$.}
    \label{fig:fig1}
\end{figure}
Intuitively, in order for User $k^\prime$ to recover the  input sum $\sum_{i\in [K] \bksl \{k^\prime\} }W_i=W_k +  \sum_{[K]\bksl \{k^\prime,k\} }W_i$, which includes $W_k$ as a summand since $ k\in [K] \bksl \{k^\prime\} $, the input $W_k$ must be conveyed through the \msg $X_k$.  Because $W_k$ is only available at User $k$, it must be fully encoded in $X_k$. Consequently, the conditional entropy of $X_k$ -- given all other users' inputs and keys---must be at least $H(W_k)$. More formally, we have
\begin{subequations}
\label{eq: proof H(Xk|(Wi,Zi) all other i), lemma}
\begin{align}
& H\left(X_k| \{W_i, Z_i\}_{i\in [K]\bksl \{k\}  } \right) \notag\\
&\ge  I\left(X_k; \sum_{i\in  [K]\bksl \{k^\prime\} } W_i \bigg|   \{W_i, Z_i\}_{i\in [K]\bksl \{k\}}         \right) \\
& = H\left(\sum_{i\in  [K]\bksl \{k^\prime\} } W_i \bigg|   \{W_i, Z_i\}_{i\in [K]\bksl \{k\}} \right)  \notag\\
& \qquad -  H\left(\sum_{i\in  [K]\bksl \{k^\prime\} } W_i \bigg|   \{W_i, Z_i\}_{i\in [K]\bksl \{k\}}, X_k \right)\\
& \overset{(\ref{eq:H(Xk|Wk,Zk)=0})}{=} H\left(W_k |   \{W_i, Z_i\}_{i\in [K]\bksl \{k\}} \right)  \notag\\
&  -  H\left(\sum_{i\in  [K]\bksl \{k^\prime\} } W_i \bigg|   \{W_i, Z_i\}_{i\in [K]\bksl \{k\}}, X_k, \{X_i\}_{i\in [K]\bksl \{k\}} \right)\label{eq: step 0, proof H(Xk|(Wi,Zi) all other i), lemma}\\
& \overset{(\ref{eq: indep. between keys and inputs})}{\ge}   H\left(W_k |   \{W_i\}_{i\in [K]\bksl \{k\}} \right)\notag\\
& \qquad  - \underbrace{H\left(\sum_{i\in  [K]\bksl \{k^\prime\} } W_i \bigg| \{X_i\}_{i\in [K]} \right)}_{ \overset{(\ref{eq:recovery constraint})}{=0} }\label{eq: step 1, proof H(Xk|(Wi,Zi) all other i), lemma}\\
& \overset{(\ref{eq:input independence})}{=}    H\left(W_k \right)\label{eq: step 2, proof H(Xk|(Wi,Zi) all other i), lemma}=L,
\end{align}
\end{subequations}
where (\ref{eq: step 0, proof H(Xk|(Wi,Zi) all other i), lemma}) is  due  to the fact that $k \in  [K]\bksl \{k^\prime\}$, and also $X_k$ is a deterministic function of $W_k$ and $Z_k,\forall  k\in[K]$ (see (\ref{eq:H(Xk|Wk,Zk)=0})). The first term in (\ref{eq: step 1, proof H(Xk|(Wi,Zi) all other i), lemma}) is due to the \indepce between the keys and inputs (see (\ref{eq: indep. between keys and inputs})),  while  the second term is due to the recovery constraint  at User $k^\prime$ (see (\ref{eq:recovery constraint})). Moreover, (\ref{eq: step 2, proof H(Xk|(Wi,Zi) all other i), lemma}) is due the \indepce of the inputs (see (\ref{eq:input independence})). Due to symmetry, (\ref{eq: proof H(Xk|(Wi,Zi) all other i), lemma}) holds for every $k\in [K]$. 
\Aar, the proof of Lemma~\ref{lemma: H(Xk|(Wi,Zi) all other i)} is complete. 
It is worth mentioning that this lemma arises solely from the recovery constraint imposed on the users, and is independent of the security requirement (\ref{eq:security constraint}). \Thf, a straightforward cut-set bound between Users $(k,k^\prime)$ suffices to derive the lower bound of $L$.
\end{IEEEproof}

Let $\Tc_{(k)} \subset [K]\bksl \{k\}$, with $|\Tc_{(k)} |\le K-3$, denote a subset of users that may collude with User $k$. We define its complement---relative to  the full set  $[K]\bksl\{k\}$---as $ \overline{\Tc}_{(k)}  \eqdef \left([K]\bksl\{k\}\right)\bksl \Tc_{(k)} $.
{A direct implication of Lemma~\ref{lemma: H(Xk|(Wi,Zi) all other i)} is the following corollary, which states that the joint entropy of the \msgs $\{X_i\}_{i\in \overline{\Tc}_{(k)} }$, when conditioned on the inputs and keys of the colluding users $\Tc_{(k)}$ and User $k$, is at least $|\overline{\Tc}_{(k)}|L$.}

\begin{corollary}
\label{corollary1}
\emph{For any colluding user set $\Tc_{(k)} \subset [K]\bksl \{k\}$ and its complement $\overline{\Tc}_{(k)}$,  it holds that}
\be
\label{eq:corollary1}
H\left( \{X_i\}_{i\in \overline{\Tc}_{(k)} } \big| \{W_i,Z_i\}_{i\in\Tc_{(k)}}, W_k,Z_k       \right) \ge |\overline{\Tc}_{(k)} |L.
\ee 
\end{corollary}
\begin{IEEEproof}
Suppose $\overline{\Tc}_{(k)}=\big\{k_1, \cdots, k_{|\overline{\Tc}_{(k)}|}\big\}$. By the chain rule of entropy, we have
\begin{subequations}
\label{eq:proof corollary1}
\begin{align}
& H\left( \{X_i\}_{i\in \overline{\Tc}_{(k)} } \big| \{W_i,Z_i\}_{i\in\Tc_{(k)}}, W_k,Z_k       \right )\notag\\
& =  \sum_{j=1}^{|\overline{\Tc}_{(k)}|} H\left( X_{k_j} \big| \{W_i,Z_i\}_{i\in\Tc_{(k)}\cup\{k\}  }, \{X_{k_i}\}_{i\in[1:j-1]}      \right)  \notag   \\
& \ge  \sum_{j=1}^{|\overline{\Tc}_{(k)}|} H\Big( X_{k_j} \big| \{W_i,Z_i\}_{i\in\Tc_{(k)}\cup\{k\}v\cup\{k_1,\cdots, k_{j-1}\}  },\notag\\
&\qquad  \qquad \quad \{X_{k_i}\}_{i\in[1:j-1]}      \Big)\label{eq:step0,proof corollary1}\\  
& \overset{(\ref{eq:H(Xk|Wk,Zk)=0})}{=} \sum_{j=1}^{|\overline{\Tc}_{(k)}|} H\left( X_{k_j} \big| \{W_i,Z_i\}_{i\in\Tc_{(k)}\cup\{k\}\cup\{k_1,\cdots, k_{j-1}\}  }\right)\label{eq:step1,proof corollary1}\\ 
& \ge \sum_{j=1}^{|\overline{\Tc}_{(k)}|} H\left( X_{k_j} \big| \{W_i,Z_i\}_{i\in[K]\bksl \{k_j\}  }\right)\label{eq:step2,proof corollary1}\\ 
& \overset{(\ref{eq: H(Xk|(Wi,Zi) all other i), lemma})}{\ge}  |\overline{\Tc}_{(k)}|L,
\end{align}
\end{subequations}
where (\ref{eq:step1,proof corollary1}) is because $X_{k_i}$ is a deterministic function of $W_{k_i}$ and $Z_{k_i}, \forall i\in [1:j-1]$ (see (\ref{eq:H(Xk|Wk,Zk)=0})). (\ref{eq:step2,proof corollary1}) is due  to the fact  that   $ \Tc_{(k)}\cup\{k\}\cup\{k_1,\cdots, k_{j-1}\} \subseteq   [K]\bksl \{k_j\}, \forall j=1,\cdots,|\overline{\Tc}_{(k)}| $. In the last step, Lemma~\ref{lemma: H(Xk|(Wi,Zi) all other i)} is applied. 
\end{IEEEproof}

{The following lemma states that any \msg  $X_k$ must be \indep of the \corrspdg input $W_k$, conditioned on the input $W_{k^\prime}$ and key $Z_{k^\prime}$ available at any other User $k^\prime \ne k$.} This follows from the security constraint imposed at User $k^\prime$, which  requires that $W_k$ to be fully protected by $Z_k$. In other words, $X_k$ must be \indep of $W_k$,
which translates to $I(X_k; W_k|W_{k^\prime}, Z_{k^\prime})=0,\forall k\ne k^\prime$. The conditioning on $W_{k^\prime}$ and $Z_{k^\prime}$
reflects the fact that User $k^\prime$ has access to this \info, and hence it must be treated as known when evaluating the mutual information.

\begin{lemma}
\label{lemma: I(Xk;Wk|Wk',Zk')=0}
\emph{For any pair of users $(k, k^\prime)$, it holds that
\begin{align}
\label{eq:lemma: I(Xk;Wk|Wk',Zk')=0}
I\left(X_k; W_k |W_{k^\prime}, Z_{k^\prime}     \right)=0.
\end{align}
}
\end{lemma}
\begin{IEEEproof}
First, by choosing $\Tc=\emptyset$ in (\ref{eq:security constraint}), we have
\be
\label{eq: security with T=emptyset, lemma}
I\left( \{X_i\}_{i \in [K]\bksl \{k^\prime\}}; \{W_i\}_{i \in [K]\bksl \{k^\prime\}} \Big|
 \sum_{i=1}^K W_i, 
 W_{k^\prime} , Z_{k^\prime}  \right)=0.
\ee 
We have
\begin{subequations}
\label{eq:proof I(Xk;Wk|Wk',Zk')=0}
\begin{align}
& I\left(X_k; W_k |W_{k^\prime}, Z_{k^\prime}     \right) \notag\\
& \le  I\left(X_k, \sum_{i=1}^KW_i; W_k \Big|W_{k^\prime}, Z_{k^\prime}     \right)\\
& =   I\left(\sum_{i=1}^KW_i; W_k \Big|W_{k^\prime}, Z_{k^\prime}\right)\notag\\
& \quad + I\left(X_k; W_k \Big|W_{k^\prime}, Z_{k^\prime}, \sum_{i=1}^KW_i     \right)\\
& \le   I\left(\sum_{i=1}^KW_i; W_k \Big|W_{k^\prime}, Z_{k^\prime}\right)\notag\\
& \quad + \underbrace{I\left( \{ X_i\}_{i\in [K]\bksl \{k^\prime\} };\{ W_i\}_{i\in [K]\bksl \{k^\prime\} } \Big|W_{k^\prime}, Z_{k^\prime}, \sum_{i=1}^KW_i     \right)}_{ \overset{(\ref{eq: security with T=emptyset, lemma})}{=}0 } \\
& \overset{(\ref{eq: indep. between keys and inputs})}{=}   I\left(\sum_{i=1}^KW_i; W_k \Big|W_{k^\prime}\right)\label{eq:step 0, proof I(Xk;Wk|Wk',Zk')=0}\\
& \overset{(\ref{eq:input independence})}{=}   I\left(\sum_{i\in[K]\bksl \{k^\prime\}  }^KW_i; W_k \right)\label{eq:step 1, proof I(Xk;Wk|Wk',Zk')=0}\\
& = H\left(\sum_{i\in[K]\bksl \{k^\prime\}  }^KW_i \right)- H\left(\sum_{i\in[K]\bksl \{k^\prime,k\}  }^KW_i \right)\label{eq:step 2, proof I(Xk;Wk|Wk',Zk')=0}\\
& \overset{(\ref{eq:input independence})}{=} L-L=0,\label{eq:step 3, proof I(Xk;Wk|Wk',Zk')=0}
\end{align}
\end{subequations}
where (\ref{eq:step 0, proof I(Xk;Wk|Wk',Zk')=0})
and (\ref{eq:step 1, proof I(Xk;Wk|Wk',Zk')=0}) are due to the \indepce between the inputs and keys (see (\ref{eq: indep. between keys and inputs})), and the \indepce of  the inputs (see (\ref{eq:input independence})), \resp. In (\ref{eq:step 2, proof I(Xk;Wk|Wk',Zk')=0}), we have $H\big(\sum_{i\in[K]\bksl \{k^\prime,k\}  }^KW_i \big)= L$ because there are at least $K\ge 3$ users by assumption (and also the  uniformity of the inputs). Moreover,
(\ref{eq:step 3, proof I(Xk;Wk|Wk',Zk')=0}) is due to the uniformity and \indepce of the inputs. Since \muinfo is non-negative, we conclude that $I\left(X_k; W_k |W_{k^\prime}, Z_{k^\prime}     \right)=0$,  completing the proof of Lemma~\ref{lemma: I(Xk;Wk|Wk',Zk')=0}.
\end{IEEEproof}

\begin{lemma}
\label{lemma: I((Xi)_Tkc;(Wi)_Tkc|Wk,Zk,(Wi,Zi)_Tk)=L}
\emph{
For any $k\in [K]$, any colluding user set $\Tc_{(k)} \subset [K]\bksl \{k\}$ and its complement $\overline{\Tc}_{(k)}$, the following equality holds:}
\begin{align}
\label{eq:I((Xi)_Tkc;(Wi)_Tkc|Wk,Zk,(Wi,Zi)_Tk)=L, lemma}
 & I\left(\{X_i\}_{i\in \overline{\Tc}_{(k)}}; 
\{W_i\}_{i\in \overline{\Tc}_{(k)}} \big|\{W_i,Z_i\}_{i\in \Tc_{(k)}}, W_k,Z_k\right)=L.
\end{align} 
\end{lemma}
\begin{IEEEproof}
We have
\begin{subequations}
\label{eq:proof I((Xi)_Tkc;(Wi)_Tkc|Wk,Zk,(Wi,Zi)_Tk)=L, lemma}
\begin{align}
& I\left(\{X_i\}_{i\in \overline{\Tc}_{(k)}}; 
\{W_i\}_{i\in \overline{\Tc}_{(k)}} \big|\{W_i,Z_i\}_{i\in \Tc_{(k)}   }, W_k,Z_k\right) \notag\\
&= I\left(\{X_i\}_{i\in \overline{\Tc}_{(k)}}; 
\{W_i\}_{i\in \overline{\Tc}_{(k)}}, \sum_{i\in \overline{\Tc}_{(k)}   } W_i \bigg|\{W_i,Z_i\}_{i\in \Tc_{(k)} \cup  \{k\} }\right)\\
& = I\left(\{X_i\}_{i\in \overline{\Tc}_{(k)}}; 
\sum_{i\in \overline{\Tc}_{(k)}   } W_i \bigg|\{W_i,Z_i\}_{i\in \Tc_{(k)} \cup  \{k\} }\right)\notag\\
& + I\left(\{X_i\}_{i\in \overline{\Tc}_{(k)}}; 
\{W_i\}_{i\in \overline{\Tc}_{(k)}} \bigg|\{W_i,Z_i\}_{i\in \Tc_{(k)}\cup  \{k\} }, \sum_{i\in \overline{\Tc}_{(k)}   } W_i\right).\label{eq:step 0, proof I((Xi)_Tkc;(Wi)_Tkc|Wk,Zk,(Wi,Zi)_Tk)=L, lemma}
\end{align}
\end{subequations}
We calculate the two terms in (\ref{eq:step 0, proof I((Xi)_Tkc;(Wi)_Tkc|Wk,Zk,(Wi,Zi)_Tk)=L, lemma}) \resp. First,
\begin{subequations}
\label{eq:part1,proof I((Xi)_Tkc;(Wi)_Tkc|Wk,Zk,(Wi,Zi)_Tk)=L, lemma}
\begin{align}
& I\left(\{X_i\}_{i\in \overline{\Tc}_{(k)}}; 
\sum_{i\in \overline{\Tc}_{(k)}   } W_i \bigg|\{W_i,Z_i\}_{i\in \Tc_{(k)} \cup  \{k\} }\right)\notag\\
& = H\left( \sum_{i\in \overline{\Tc}_{(k)}   } W_i \bigg|\{W_i,Z_i\}_{i\in \Tc_{(k)} \cup  \{k\} }\right) \notag\\
& \quad - H\left( \sum_{i\in \overline{\Tc}_{(k)}   } W_i \bigg|\{W_i,Z_i\}_{i\in \Tc_{(k)}\cup  \{k\} }, \{X_i\}_{i\in \overline{\Tc}_{(k)}}\right)\\
&  \overset{(\ref{eq:input independence}),(\ref{eq: indep. between keys and inputs}),(\ref{eq:H(Xk|Wk,Zk)=0})}{=} H\left( \sum_{i\in \overline{\Tc}_{(k)}   } W_i \right) \notag\\
& - H\left( \sum_{i\in [K]   } W_i \bigg|\{W_i,Z_i\}_{i\in \Tc_{(k)} \cup  \{k\} }, \{X_i\}_{i\in [K]\bksl\{k\} }\right)\label{eq:step0,part1,proof I((Xi)_Tkc;(Wi)_Tkc|Wk,Zk,(Wi,Zi)_Tk)=L, lemma}\\
& \overset{(\ref{eq:recovery constraint})}{=}   H\left( \sum_{i\in \overline{\Tc}_{(k)}   } W_i \right),
\end{align}
\end{subequations}
where the second term in (\ref{eq:step0,part1,proof I((Xi)_Tkc;(Wi)_Tkc|Wk,Zk,(Wi,Zi)_Tk)=L, lemma}) is equal to zero because $
H\left( \sum_{i\in [K]   } W_i|\{W_i,Z_i\}_{i\in \Tc_{(k)} \cup  \{k\} }, \{X_i\}_{i\in [K]\bksl\{k\} }\right)\le H\left( \sum_{i\in [K]   } W_i |W_k,Z_k, \{X_i\}_{i\in [K]\bksl\{k\} }\right)=0
$ according to the recovery constraint (\ref{eq:recovery constraint}) imposed on User $k$.

Recall that $\Cc_{\Sc}=\{W_i,Z_i\}_{i\in \Sc} $ represents the collection of inputs and keys at the users $\Sc$. We have
\begin{subequations}
\label{eq:part2,proof I((Xi)_Tkc;(Wi)_Tkc|Wk,Zk,(Wi,Zi)_Tk)=L, lemma}
\begin{align}
& I\left(\{X_i\}_{i\in \overline{\Tc}_{(k)}}; 
\{W_i\}_{i\in \overline{\Tc}_{(k)}} \bigg|\{W_i,Z_i\}_{i\in \Tc_{(k)} \cup  \{k\} }, \sum_{i\in \overline{\Tc}_{(k)}   } W_i\right)\notag\\
& \le 
I\left(\{X_i\}_{i\in[K]\bksl \{k\} }; 
\{W_i\}_{i\in[K]\bksl \{k\} } \bigg|\Cc_{\Tc_{(k)} \cup  \{k\}}, \sum_{i\in \overline{\Tc}_{(k)}   } W_i\right)\label{eq:step0,part2,proof I((Xi)_Tkc;(Wi)_Tkc|Wk,Zk,(Wi,Zi)_Tk)=L, lemma}\\
& =
I\left(\{X_i\}_{i\in[K]\bksl \{k\} }; 
\{W_i\}_{i\in[K]\bksl \{k\} } \bigg|\Cc_{\Tc_{(k)}\cup  \{k\}}, \sum_{i\in [K]   } W_i\right)\label{eq:step1,part2,proof I((Xi)_Tkc;(Wi)_Tkc|Wk,Zk,(Wi,Zi)_Tk)=L, lemma}\\
& \overset{(\ref{eq:security constraint})}{=}0,
\end{align}
\end{subequations}
where (\ref{eq:step0,part2,proof I((Xi)_Tkc;(Wi)_Tkc|Wk,Zk,(Wi,Zi)_Tk)=L, lemma}) is because $\overline{\Tc}_{(k)} \subseteq [K]\bksl \{k\}$, and the last  step is due to the security constraint of User $k$. Since \muinfo is non-negative, the second term in (\ref{eq:step 0, proof I((Xi)_Tkc;(Wi)_Tkc|Wk,Zk,(Wi,Zi)_Tk)=L, lemma}) is equal to zero.
Plugging (\ref{eq:part1,proof I((Xi)_Tkc;(Wi)_Tkc|Wk,Zk,(Wi,Zi)_Tk)=L, lemma}) and (\ref{eq:part2,proof I((Xi)_Tkc;(Wi)_Tkc|Wk,Zk,(Wi,Zi)_Tk)=L, lemma}) into (\ref{eq:proof I((Xi)_Tkc;(Wi)_Tkc|Wk,Zk,(Wi,Zi)_Tk)=L, lemma}), we have
\begin{subequations}
\begin{align}
 & I\left(\{X_i\}_{i\in \overline{\Tc}_{(k)}}; 
\{W_i\}_{i\in \overline{\Tc}_{(k)}} \big|\{W_i,Z_i\}_{i\in \Tc_{(k)}}, W_k,Z_k\right)\notag\\
 & = H\left( \sum_{i\in \overline{\Tc}_{(k)}   } W_i \right) \overset{(\ref{eq:input independence})}{=} L,
\end{align}
\end{subequations}
where the last step follows from the the  uniformity and the \indepce of the inputs. Note that $|\overline{\Tc}_{(k)}|\ge 2$ because $|\Tc_{(k)}|\le K-3$  for \secagg to be feasible. 
\Aar, Lemma~\ref{lemma: I((Xi)_Tkc;(Wi)_Tkc|Wk,Zk,(Wi,Zi)_Tk)=L} is proved.
\end{IEEEproof}

{Lemma~\ref{lemma: I((Xi)_Tkc;(Wi)_Tkc|Wk,Zk,(Wi,Zi)_Tk)=L} implies that, from  the perspective of any User $k$, even if it colludes with  a  subset of users $\Tc_{(k)}\subset[K]\bksl \{k\}$, the only \info that can be  inferred about the inputs $\{W_i\}_{i\in \overline{\Tc}_{(k)}} $ is their sum.} This reflects the core security requirement imposed on User $k$: it must not learn anything beyond the sum of inputs and the information already available through collusion, regardless of which subset of users it colludes with.

We also have the following lemma, which asserts  that the joint entropy of the  \indiv  keys in any $\overline{\Tc}_{(k)}$, conditioned on all other keys in $\Tc_{(k)}\cup \{k\}$, is at least $(|\overline{\Tc}_{(k)}|-1)L$ for any User $k\in [K]$.
\begin{lemma}
\label{lemma:H((Zi)_Tkc|(Zi,Wi)_Tk, Wk,Zk)>=(K-|Tkc|-2)L}
\emph{
For any set of colluding users $\Tc_{(k)}\in [K]\bksl\{k\}$ and its complement $\overline{\Tc}_{(k)}=( [K]\bksl\{k\})\bksl \Tc_{(k)}$, it holds that}
\be
\label{eq:H((Zi)_Tkc|(Zi,Wi)_Tk, Wk,Zk)>=(K-|Tkc|-2)L, lemma}
H\left( \{Z_i\}_{ i\in\overline{\Tc}_{(k)}  } \big| \{Z_i\}_{i\in \Tc_{(k)}},Z_k   \right)
\ge \left(K-2-|\Tc_{(k)}|\right)L.
\ee
\end{lemma}

\begin{IEEEproof}
We have
\begin{subequations}
\label{eq:proof,H((Zi)_Tkc|(Zi,Wi)_Tk, Wk,Zk)>=(K-|Tkc|-2)L, lemma}
\begin{align}
& H\left( \{Z_i\}_{ i\in\overline{\Tc}_{(k)}  } \big| \{Z_i\}_{i\in \Tc_{(k)}},Z_k   \right)\notag\\
& \overset{(\ref{eq: indep. between keys and inputs})}{=}
H\left( \{Z_i\}_{ i\in\overline{\Tc}_{(k)}  } \big|\{W_i\}_{ i\in\overline{\Tc}_{(k)}  } , \{Z_i\}_{i\in \Tc_{(k)}},W_k,Z_k 
 \right)\label{eq:step0,proof,H((Zi)_Tkc|(Zi,Wi)_Tk, Wk,Zk)>=(K-|Tkc|-2)L, lemma}\\
& \ge I\Big( \{Z_i\}_{ i\in\overline{\Tc}_{(k)}  }; \{X_i\}_{ i\in\overline{\Tc}_{(k)}  }  \big|\{W_i\}_{ i\in\overline{\Tc}_{(k)}  } , \{Z_i\}_{i\in \Tc_{(k)}},\notag\\
&\qquad\quad W_k,Z_k \Big)\\
& = H\left( \{X_i\}_{ i\in\overline{\Tc}_{(k)}  }  \big|\{W_i\}_{ i\in\overline{\Tc}_{(k)}  } , \{Z_i\}_{i\in \Tc_{(k)}}, W_k,Z_k \right)\notag\\
& - \underbrace{H\left( \{X_i\}_{ i\in\overline{\Tc}_{(k)}  }  \big|\{W_i,Z_i\}_{ i\in\overline{\Tc}_{(k)}  } , \{Z_i\}_{i\in \Tc_{(k)}}, W_k,Z_k \right)}_{\overset{(\ref{eq:H(Xk|Wk,Zk)=0})}{=}0  }\\
& \ge 
H\left( \{X_i\}_{ i\in\overline{\Tc}_{(k)}  }  \big|\{W_i\}_{ i\in\overline{\Tc}_{(k)}  } , \{W_i,Z_i\}_{i\in \Tc_{(k)}}, W_k,Z_k \right)\\
& = 
H\left( \{X_i\}_{ i\in\overline{\Tc}_{(k)}  }  \big| \{W_i,Z_i\}_{i\in \Tc_{(k)}}, W_k,Z_k \right)  \notag\\
&- I\left( \{X_i\}_{ i\in\overline{\Tc}_{(k)}  }; \{W_i\}_{ i\in\overline{\Tc}_{(k)}  }  \big| \{W_i,Z_i\}_{i\in \Tc_{(k)}}, W_k,Z_k \right) \\
& \overset{(\ref{eq:corollary1}),(\ref{eq:I((Xi)_Tkc;(Wi)_Tkc|Wk,Zk,(Wi,Zi)_Tk)=L, lemma})}{\ge }
\left(|\overline{\Tc}_{(k)}|-1\right)L\label{eq:step1,proof,H((Zi)_Tkc|(Zi,Wi)_Tk, Wk,Zk)>=(K-|Tkc|-2)L, lemma}\\
& = \left(K-2-|\Tc_{(k)}|\right)L,
\end{align}
\end{subequations}
where  (\ref{eq:step0,proof,H((Zi)_Tkc|(Zi,Wi)_Tk, Wk,Zk)>=(K-|Tkc|-2)L, lemma}) is due to the \indepce between the inputs and keys. In (\ref{eq:step1,proof,H((Zi)_Tkc|(Zi,Wi)_Tk, Wk,Zk)>=(K-|Tkc|-2)L, lemma}), Corollary~\ref{corollary1} and Lemma~\ref{lemma: I((Xi)_Tkc;(Wi)_Tkc|Wk,Zk,(Wi,Zi)_Tk)=L} are applied. The last step is because $|\overline{\Tc}_{(k)}| +|\Tc_{(k)}|=K-1$.  \Aar,  the proof of Lemma~\ref{lemma:H((Zi)_Tkc|(Zi,Wi)_Tk, Wk,Zk)>=(K-|Tkc|-2)L} is complete. 
\end{IEEEproof}

\begin{figure}[ht]
    \centering
    \includegraphics[width=0.33\textwidth]{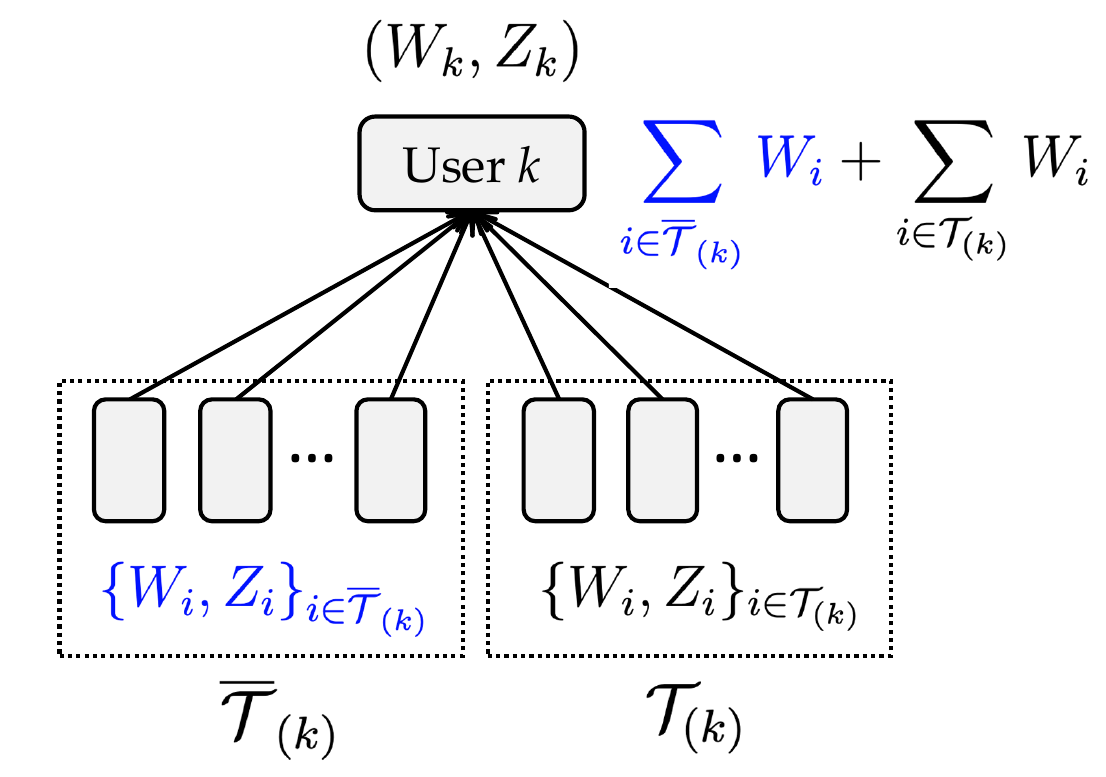}
    \vspace{-.4cm}
    \caption{\Agg at User $k$ with colluding user set $\Tc_{(k)}$ and its complement $\overline{\Tc}_{(k)}$. Since the inputs $\{W_i\}_{i\in\Tc_{(k)}}$ are already known to User $k$, it only needs to recover the sum of inputs in $\sum_{i\in \overline{\Tc}_{(k)}}W_i $ (shown in blue), while ensuring their security.}
    \label{fig:fig2}
\end{figure}

{An intuitive interpretation of Lemma~\ref{lemma:H((Zi)_Tkc|(Zi,Wi)_Tk, Wk,Zk)>=(K-|Tkc|-2)L} is provided \af.}
Let us consider the \agg at User $k$, as shown in Fig.~\ref{fig:fig2}.
Since the inputs and keys $ \{W_i,Z_i\}_{i\in \Tc_{(k)}}$ are already known to User $k$ through collusion, User $k$ only needs  to recover the sum of inputs $\sum_{i\in \overline{\Tc}_{(k)}}W_i$ of the remaining users $\overline{\Tc}_{(k)}$. To ensure the security of these inputs against User $k$, a fundamental result by Zhao and Sun~\cite{zhao2023secure} states that the users in $\overline{\Tc}_{(k)}$ must collectively hold at least $(|\overline{\Tc}_{(k)}|-1)L$ \indep key symbols. Given that User $k$  also possesses  $Z_k$ in addition to the colluded keys $\{Z_i\}_{i\in \Tc_{(k)}}$, we have 
$
H\big( \{Z_i\}_{ i\in\overline{\Tc}_{(k)}  } \big| \{Z_i\}_{i\in \Tc_{(k)}},Z_k \big)
\ge (|\overline{\Tc}_{(k)}|-1)L=(K-2-|\Tc_{(k)}|)L
$.

\subsection{Lower Bounds on \Comm and Key Rates}
\label{subsec:lower bounds on comm and key rates}
Equipped with the lemmas in the previous section, we now derive the lower bounds on the \comm and key rates. 
{\Ip, the proof of the \comm rate $\rx\ge 1$ relies on Lemma \ref{lemma: H(Xk|(Wi,Zi) all other i)} which reflects the requirement of the recovery constraint (\ref{eq:recovery constraint}). The proof of the key rates  $\rz\ge  1, \rzsigma\ge  K-1$ relies primarily on Lemmas \ref{lemma: I(Xk;Wk|Wk',Zk')=0}, \ref{lemma: I((Xi)_Tkc;(Wi)_Tkc|Wk,Zk,(Wi,Zi)_Tk)=L} and  \ref{lemma:H((Zi)_Tkc|(Zi,Wi)_Tk, Wk,Zk)>=(K-|Tkc|-2)L} which reflect the requirement of the security constraint (\ref{eq:security constraint}).} 
Since these bounds match the \achvb rates in Section~\ref{sec: ach scheme}, the optimality of the proposed scheme can be established.

\subsubsection{Proof of \Comm Rate $\rx \ge 1$}
\label{subsubsec:proof of Rx>=1}
Consider any User $k\in[K]$. 
A straightforward application of Lemma~\ref{lemma: H(Xk|(Wi,Zi) all other i)} gives the following bound on the \comm rate:
\begin{subequations}
\begin{align}
L_X \ge H(X_k) & \ge H\left(X_k |\{W_i, Z_i\}_{i\in [K]\bksl \{k\}   } \right)\overset{(\ref{eq: H(Xk|(Wi,Zi) all other i), lemma})}{\ge } L,\\
\Rightarrow \rx  & \eqdef {L_X}/{L} \ge  1.
\end{align}
\end{subequations}
{Note that this bound relies on the \iid property of the inputs.}

\subsubsection{Proof of \Indiv Key Rate $\rz \ge 1$}
\label{subsubsec:proof of Rz>=1}
Consider any pair of users $(k,k^\prime)$. We have
\begin{subequations}
\label{eq:proof of Rz>=1,converse}
\begin{align}
L_Z & \ge H(Z_k)\\
&  \ge  H\left(Z_k|W_k, W_{k^\prime}, Z_{k^\prime} \right  )\\
 & \ge I\left(Z_k; X_k|W_k, W_{k^\prime}, Z_{k^\prime}   \right)\\
 & = H\left(X_k|W_k, W_{k^\prime}, Z_{k^\prime} \right) - \underbrace{ H\left(X_k|W_k, Z_k,W_{k^\prime}, Z_{k^\prime} \right)}_{\overset{(\ref{eq:H(Xk|Wk,Zk)=0})}{=} 0   }\notag\\
& = H\left(X_k|W_{k^\prime}, Z_{k^\prime} \right) - \underbrace{ I\left(X_k;W_k|W_{k^\prime}, Z_{k^\prime} \right)}_{\overset{(\ref{eq:lemma: I(Xk;Wk|Wk',Zk')=0})}{=}0 } \label{eq:step0,proof of Rz>=1,converse}\\
& \ge  H\left(X_k|\{W_i,Z_i\}_{ i\in  [K]\bksl\{k\}  } \right)\label{eq:step1,proof of Rz>=1,converse}\\
& \overset{(\ref{eq: H(Xk|(Wi,Zi) all other i), lemma})}{\ge}L \Rightarrow  \rz  \eqdef {L_Z}/{L}\ge 1,\label{eq:step2,proof of Rz>=1,converse}
\end{align}
\end{subequations}
where {Lemma~\ref{lemma: I(Xk;Wk|Wk',Zk')=0}} is applied in (\ref{eq:step0,proof of Rz>=1,converse}). (\ref{eq:step1,proof of Rz>=1,converse}) is because $k^\prime \in[K]\bksl\{k\} $ since $k\ne k^\prime$. Moreover, Lemma~\ref{lemma: H(Xk|(Wi,Zi) all other i)} is applied in (\ref{eq:step2,proof of Rz>=1,converse}).

\subsubsection{Proof of Source Key Rate $\rzsigma \ge K-1$}
\label{subsubsec:proof of Rzsigma>=1} 
Consider any $k\in[K]$. By setting the colluding user set as $\Tc_{(k)}= \emptyset$ (so that $\overline{\Tc}_{(k)}=[K]\bksl\{k\}$) in Lemma~\ref{lemma:H((Zi)_Tkc|(Zi,Wi)_Tk, Wk,Zk)>=(K-|Tkc|-2)L}, we obtain
\be
\label{eq:eq1,proof of Rzsigma>=K-1}
H\left( \{Z_i\}_{i\in [K]\bksl\{k\}} |Z_k   \right) \ge(K-2)L.
\ee 
Then 
\begin{subequations}
\label{eq:proof of Rzsigma>=K-1}
\begin{align}
\lzsigma \ge H\left(\zsigma  \right)
 &  \overset{(\ref{eq: H(Z1,...,ZK|Zsigma)=0})}{=}H\left(\zsigma  \right) + H\left(Z_{1:K}|\zsigma  \right)\label{eq:step0,proof of Rzsigma>=K-1} \\
 & = H\left(Z_{1:K}, \zsigma\right)\\
 & \ge H(Z_{1:K})\\
 & = H(Z_k) + H\left( \{Z_i\}_{i\in [K]\bksl\{k\}} |Z_k   \right)\\
& \overset{(\ref{eq:proof of Rz>=1,converse}),(\ref{eq:eq1,proof of Rzsigma>=K-1})}{\ge }
L + (K-2)L\\
&= (K-1)L,\\
\Rightarrow & \rzsigma  \eqdef {\lzsigma}/{L} \ge K-1,
\end{align}
\end{subequations}
where (\ref{eq:step0,proof of Rzsigma>=K-1}) is because the \indiv keys are generated from the source key (see (\ref{eq: H(Z1,...,ZK|Zsigma)=0})). \Aar, we proved  $\rzsigma \ge K-1$.

\section{Conclusion and Future Directions}
\label{sec:conclusion & future directions}
In this work, we studied the problem of collusion-resilient \dsa (DSA) from an \itic perspective. Specifically, users in a  fully-connected network aim to simultaneously compute the sum of the private inputs of all users through collaborative \comm.
The security constraint ensures that besides the sum of the  inputs, no \info about  the inputs should be revealed to any user, even in the presence of collusion among users. 
We fully characterized the optimal rate region, which specifies the minimally \achvb \comm, \indiv and source key rates. A major finding is that
the decentralized network structure enables optimal reuse of the dual roles of each individual key---serving both as encryption for the corresponding input and as an antidote for recovering the input sum.

Several future directions can be explored: 
{1) DSA over arbitrary or sparse network topologies, where each user connects to a varying number of neighbors and aims to recover the sum of their inputs. In this case, the optimal source key rate may depend on the largest node degree,  and
the required individual key rate may differ across users, necessitating a characterization of the optimal sequence of key rates}; 
2) DSA with user dropout resilience, where the aggregation scheme must ensure that the surviving users can robustly recover the sum of inputs even in the presence of user dropout; 
3) DSA with groupwise keys, where each subset of users shares an independent key (termed groupwise key). These keys can be generated through interactive key agreement protocols over user-to-user communication, thereby removing the need for a centralized key distribution server.

\appendices

\section{The Baseline Scheme}
\label{sec:baseline scheme,appendix}
{Motivated by the centralized \secagg scheme proposed by Zhao and Sun~\cite{zhao2023secure},
a straightforward baseline for DSA is to apply the Zhao-Sun scheme repeatedly over $K$ rounds.
In each round, one user is designated as a \tit{virtual} server that aggregates the inputs of the remaining users.
Specifically,  in the \kth  round, the goal is to compute the input sum $\sum_{i\in [K]\bksl \{k\}} W_i $ at User $k\in  [K]$. 
To this end, a collection of  $K$ secret keys $\big\{N_{1}^{(k)}, \cdots, N_K^{(k)}\big\}$ is generated, where $N_{k}^{(k)}$ is assigned (by the key server) to User $k, \forall k\in  [K]$. 
These keys satisfy 1) $\big\{N_i^{(k)}: i \in \Kexclk \big\}$  are $K-1$ \iid random symbols from $\mbb{F}_q$, and 2) $N_k^{(k)} = - \sum_{i\in \Kexclk  }  N_i^{(k)}  $  so that the $K$ keys have a zero sum.
Each User $i \in [K]\bksl \{k\}$ broadcasts $X_i^{(k)}=W_i+ N_i^{(k)}$ to other users. From the received \msgs $\big\{X_i^{(k)}: i\in \Kexclk     \big\}$, User $k$ can decode $\sum_{i\in \Kexclk}W_i$ by adding the \msgs with its own secret key:
$\sum_{i\in \Kexclk}X_i + N_k^{(k)}   =\sum_{i\in \Kexclk}W_i + \sum_{i=1}^K N_i^{(k)} = \sum_{i\in \Kexclk}W_i$. Note that User $k$ does not transmit in the \kth round. 
By repeating the above process $K$ rounds, each 
round recovering the aggregate sum for one user, the input sum can be obtained at all $K$ users. To avoid  input leakage across rounds, an \indep collection of keys is used in each round; that is, the key sets $\big\{N_i^{(1)}\big\}_{i\in [K]},\cdots, \big\{N_i^{(K)}\big\}_{i\in [K]}   $ are mutually \indep. 

\tbf{Security.}
The security of this baseline scheme follows from the security of the aggregate in each individual round, together with the independence of the secret keys used across different rounds. We briefly explain the former  \af. 
In the \kth round, due to the zero-sum property of the keys and the fact that any  $K-1$ keys are \indep,  the only way for User $k$ to eliminate the keys is to sum the $K-1$ received  \msgs together with its own key $N_k^{(k)}$. This operation yields the desired input sum, implying that User $k$ learns nothing beyond the input sum.
Moreover, each user $k'\ne k$   receives $K-2$ \msgs $\big\{X_i^{(k)}: i\in [K]\bksl\{k,k'\}   \big\}$, each protected by an \indep key that is also \indep of User $k'$'s key $N_{k'}^{(k)}$. Consequently, User $k'$ cannot decode any input, which establishes the security of the \kth round.

\tbf{Rates.}
Since in each round $K-1$ \msgs are sent, each user is assigned one key symbol and all users collectively hold  $K-1$ \indep  key symbols, the baseline scheme achieves $\rx=K(K-1), \rz=K, \rzsigma=K(K-1)$. Compared with the optimal DSA rates characterized in \Thm~\ref{thm:main result}, the \comm, \indiv and source key rates are inflated by factors of $K-1,K$, and $K$, \resp.
}

\bibliographystyle{IEEEtran}
\bibliography{references_secagg.bib}

\end{document}